\documentclass[12 pt]{article}
\usepackage{graphicx}
\usepackage{dcolumn}
\usepackage{amsmath}
\usepackage{amsfonts}
\usepackage{amssymb}
\usepackage{latexsym}
\begin{document}
\begin{center}
{\bf{TOPICS IN NONCOMMUTATIVE GEOMETRY INSPIRED PHYSICS}}
\end{center}
\begin{center}
\vskip 1cm Rabin Banerjee$^1$ , Biswajit Chakraborty$^1$, Subir
Ghosh$^2$, Pradip Mukherjee$^3$, Saurav Samanta$^{4}$ \vskip .5cm
{$^1$S. N. Bose National Centre for Basic Sciences,
      JD Block, Sector III, Salt Lake, Kolkata-700098, India}\\
{$^2$Physics and Applied Mathematics Unit, Indian
Statistical Institute,  Kolkata-700108, India}\\

{$^3$Presidency College, 86/1 College Street, Kolkata-700073, West-Bengal, India}\\
{$^4$Narasinha Dutt College,
129, Belilious Road, Howrah-711101, India}\\
\end{center}
\begin{abstract}
In this review article we discuss some of the applications  of
noncommutative geometry  in physics that are of recent interest,
such as noncommutative many-body systems, noncommutative extension
of Special Theory of Relativity kinematics, twisted gauge theories
and noncommutative gravity.
\end{abstract}
\newpage
{\bf{\begin{large}Table of Contents \end{large}:}}\\
\vskip .2cm
 \textbf{1. Introduction {\phantom {xxxxxxxxxxxxxxxxxxxxxxxxxxxxxxxxxxxxxxxxxx}}                                  3}\\
\vskip .1cm
 \textbf{2. Noncommutative Theories, Symmetries and Their Implications \,\,\,5}\\
2.1 Elementary Idea About Moyal Star Product in 2D and the Landau Problem \,\,5\\
2.2 Problem Regarding Poincare/Galilean symmetry and its restoration {\phantom {xxx}} \quad\quad 6\\
2.3 Operatorial Approach {\phantom {xxxxxxxxxxxxxxxxxxxxxxxxxxxxxxxxxxxxxxxxxxxxx}}10\\
\vskip .1cm
\textbf{3. Doubly Special Relativity and Noncommutativity {\phantom {xxxxxxxxxxx}}\,\,\,11}\\
3.1 DSR Phase Space {\phantom {xxxxxxxxxxxxxxxxxxxxxxxxxxxxxxxxxxxxxxxxxxxxxxxxx}}12\\
3.2 Canonical Variables {\phantom {xxxxxxxxxxxxxxxxxxxxxxxxxxxxxxxxxxxxxxxxxxx}}\quad\quad\,\,15\\
3.3 Deformed Symmetry Generators {\phantom {xxxxxxxxxxxxxxxxxxxxxxxxxxxxxxxxxxxxx}}16\\
3.4 Lagrangian for $\kappa $-Particle {\phantom {xxxxxxxxxxxxxxxxxxxxxxxxxxxxxxxxxxxxxxxxxxx}}17\\
\vskip .1cm
\textbf{4. Deformed and Twisted Gauge Symmetry \\{\phantom{xxxx}} in Noncommutative Field Theory {\phantom {xxxxxxxxxxxxxxxxxxxxxxxxxx}}18}\\
4.1 Lagrangian Analysis {\phantom {xxxxxxxxxxxxxxxxxxxxxxxxxxxxxxxxxxxxxxxxxxxxxxx}}\,19\\
4.2 Hamiltonian Analysis {\phantom {xxxxxxxxxxxxxxxxxxxxxxxxxxxxxxxxxxxxxxxxxxxxxx}}\,\,26\\
\vskip .1cm
\textbf{5. Noncommutative Gravity and Black Hole Physics {\phantom {xxxxxxxxxxxx}}\,30}\\
5.1 Lie Algebraic Noncommutative Gravity {\phantom {xxxxxxxxxxxxxxxxxxxxxxxxxxxxxxx}}\,\,31\\
5.2 Noncommutativity Inspired Black Hole Physics {\phantom {xxxxxxxxxxxxxxxxxxxxxxxx}}\,\,\,34\\
5.3 Coherent State Based Approach to Noncommutative Black Hole Physics {\phantom {xxxx}}36\\
\vskip .1cm
\textbf{6 Concluding Remarks {\phantom {xxxxxxxxxxxxxxxxxxxxxxxxxxxxxxxxxxxx}}\,\,\,42}\\
\vskip .1cm
\textbf{Acknowledgments {\phantom {xxxxxxxxxxxxxxxxxxxxxxxxxxxxxxxxxxxxxxxx}}\,\,\,43}
\newpage
\section{Introduction}
Preparing a review article on Non-Commutative (NC) geometry in
physics turns out to be a daunting task mainly because of its
great impact in diverse areas of modern physics such as Quantum
Mechanics, High Energy, Gravity and Condensed Matter. There are
excellent reviews \cite{szabo} that cover much of the earlier
work. In the present article we will concentrate on specific
topics broadly falling in each of the above branches.

NC spacetime was first  introduced by Snyder \cite{sn},
\begin{equation}
 [x_\mu ,x_\nu ]=i\theta _{\mu\nu}(x,p)=i\theta (x_\mu p_\nu - x_\nu p_\mu)
\label{01}
\end{equation}
in an attempt to introduce a short distance cutoff (the NC
parameter) in a Lorentz covariant way that can cure the
divergences in relativistic Quantum Field Theory (QFT). In
(\ref{01}) $x_\mu,p_\mu $ are the coordinate and momentum
variables and $\theta $ is the NC parameter. However it did not
become popular mainly due to its inability to address the
radiative corrections correctly that were later accounted for
successfully in the renormalization programme. Indeed, even at the
very beginning Yang \cite{yang} had hinted about such a
possibility. The remarkable rise in the interest in NC physics
started with the seminal paper by Seiberg and Witten \cite{sw}
where it was established that in certain low energy limits  String
Theory can be formulated as an effective  QFT in NC spacetime,
\begin{equation}
 [x_\mu ,x_\nu ]=i\theta _{\mu\nu}.
\label{02}
\end{equation}
Still earlier, noncommutative spacetime was also derived from open string boundary conditions in \cite{shei}.
In (\ref{02}) $\theta _{\mu\nu}$ is generally taken as an
antisymmetric {\it{constant}} tensor. A similar form of NC
structure has been shown to arise from the consideration of
boundary condition in the case of open (Super)String or membrane
moving in the background of a constant antisymmetric tensor field,
both at the classical \cite{biswa1} and quantum \cite{biswa2}
level. Notice a qualitative difference between the forms of NC in
(\ref{01}) and (\ref{02}): in the former NC is operatorial (it is
basically the angular momentum operator) or dynamical whereas in
(\ref{02}) it of a $c$-number form. In the present article we
refer to theories defined on NC spacetime as NC theories.

 Subsequently many peculiar non-perturbative features of  QFT in NC spacetime (\ref{02}),
 (these are essentially non-local effects), were revealed, such as  NC solitons without a smooth commutative limit \cite{ncsol}, formal similarity of abelian NC
 gauge theories with non-abelian gauge theories \cite{szabo}, (an example being the quantization of Chern-Simons
 coupling in abelian NC gauge theory \cite{nccs}), structural similarity of NC gauge theories with gravity \cite{ban-yang,ban-yangnew} to name a few. On the other hand, following a prescription
 (the Seiberg-Witten map) given in \cite{sw} on how to obtain NC corrections in perturbative QFT framework
 order by order in the NC parameter $\theta $ a large amount of work has appeared that studied NC extensions
 of QFT anomalies \cite{anomaly}, NC-solitons having smooth commutative limit \cite{ncsol1},  Chern-Simons
 gauge theories \cite{nccsg}, (for more recent result  valid to all orders of $\theta $ see \cite{exnc}
 that follow the general method proposed by \cite{cpm}) etc. NC generalization in Quantum Mechanical models
 such as Harmonic Oscillator \cite{ncho}, Hydrogen atom spectra \cite{nch}, NC gravitational well \cite{bertol} are geared to predict theoretical
 values of $\theta $ that can be checked experimentally. However, perturbative analysis also generates a problem, known as the Ultra violet - Infrared mixing
 \cite{uvir}, that shows that high and low energy scales of energy get entangled due to noncommutativity which can affect the renormalization programme. So far there is no conclusive experimental evidence
 for a non-vanishing $\theta $.  To name some of  important and active areas (that we have
 not touched upon in this review) we mention renormalization for NCQFT \cite{wul}, NC geometry in
 the context of spectral model of spacetime  \cite{con}, NC description of Quantum Hall Fluid \cite{fluid}, among others.

More recently, the operatorial form of NC spacetime (\ref{01}) or
a Lie algebraic form, (where $\theta _{\mu\nu}(x)$ depends only on
$x_\mu$),  is playing important roles in different contexts. An
example is in Condensed Matter Physics where planar systems
involving a perpendicular magnetic field becomes effectively
non-commutative in the lowest Landau level \cite{horv} (see Szabo in \cite{szabo})  the NC
parameter being identified with inverse of the magnetic field or
in Anyon models \cite{cnp}. The Landau levels also get
renormalized by interparticle interactions \cite{scholtz} that can have
nontrivial impact in fractional Quantum Hall Effect. In another
development it has been shown that NC phase space algebra can
influence particle dynamics directly via induced Berry curvature
effects \cite{ber} in studying many Condensed Matter phenomena
such as Anomalous Hall effect \cite{ahe,horv}, Spin Hall effect
\cite{she}, models of Graphene \cite{ber2} among others. NC
momentum algebra was used in \cite{jmm} to rederive the Dirac
Quantization Condition in a gauge invariant way and this result
has been extended in NC space in a recent work \cite{chm}.

Indeed, our review is biased in the sense that we have left out a
number of  exciting areas in NC physics some of which are
mentioned above. Our listing is, by no means, exhaustive. This
review is restrictive and presents our perspective and choice of
topics. However, in this brief {\bf{Introduction}}, we will not
dwell too much on the topics we have covered in the present review
simply because the individual chapters are reasonably
self-contained. {\bf{Section 2}} introduces the star (Moyal) product used to
define products of fields in NC spaces discusses effects of NC
space on non-relativistic many (bose/fermi) particle systems
through twisted statistics \cite{scholtz,4,5,6}. In {\bf{Section 3}} we
further develop kinematics in NC spacetime at an algebraic level
\cite{gho2} by considering the Doubly Special Relativity (DSR)
framework from the perspective of NC spacetime. The concept of
DSR, proposed by Amelino Camelia \cite{am} (as an extension to
Special Theory of Relativity), can act as the proper arena for
Quantum Gravity, since it can accommodate an {\it{observer
independent}} length scale (which might be Planck length).
{\bf{Section 4}} gives a detailed analysis of gauge symmetries in
NC gauge theories, pointing out the similarities and differences
with the conventional interpretation in ordinary (commutative
space) gauge theories \cite{rb,samanta1}. The idea goes back to
the approach proposed by Wess \cite{jWess} and by Chaichian et.al
\cite{cht,cht1} who showed that one can interpret the Lorentz (or
Poincare) symmetry violating NC theories to be {\it{invariant}}
under a Twisted Lorentz (or Poincare) symmetry{\footnote{For a
discussion on Poincare symmetry in NC field theory using Noether
prescription see \cite{l1}}}. {\bf{Section 5}} is devoted to a study
of NC effects in Gravity \cite{MS,CTZ} where Seiberg-Witten map is
utilized in computing $O(\theta )$ effects on Einstein Gravity. For an
alternative approach  based on coherent state formalism \cite{glauber} see \cite{nicolini}. We
end with concluding remarks in {\bf{Section 6}}.

\section{Noncommutative Theories, Symmetries and Their Implications}

In this section, we are going to review very briefly the Moyal
type NC space in a nonrelativistic setting touching on important
applications in Quantum Hall Effect (QHE), violation and eventual
restoration of rotational symmetry in space with dimension $D \ge
3 $. This serves the dual purpose of how dramatically this impacts
on the deformation of Bose/Fermi statistics giving rise to the
violation of Pauli's principle in one hand and on the other hand
it illustrates how the Poincare symmetry is broken and eventually
restored in straightforward manner. This restoration of Poincare
symmetry is an essential pre-requisite for the elementary
particles to be classified according to Wigner.

\subsection{Elementary Idea About Moyal Star Product in 2D and the Landau Problem}

To motivate briefly the introduction of Moyal star product
formalism let us consider the two dimensional noncommutative plane
\begin{eqnarray}
[\hat x_1, \hat x_2] &=& i\theta \label{1}
\end{eqnarray}
\begin{eqnarray}
[\hat t, \hat x_{i}] &=& 0 ; i=1,2  \nonumber
\end{eqnarray}
so that time 't' can be regarded as a usual commutating parameter.
At this stage, one can go ahead with the quantization programme by
working with the operator-valued coordinates (see sub-section 5.3)
or else demote the status of the operators '$\hat x_1$' and '$\hat
x_2$' to ordinary c-numbered valued coordinates, where the
compositions of any pair of functions thereof has to be performed
by the Moyal star product \cite{szabo}. To see this, heuristically
consider the Weyl's prescription of constructing Weyl ordered
operator $W[f(\overrightarrow{x})]$ from the c-numbered valued
function $f(\overrightarrow{x})$ of the commuting variables
$x_1,x_2$. According to this prescription, one has to just replace
$\overrightarrow{x}$ occurring in the exponent in the identity
\begin{equation}
f(\overrightarrow{x}) = \frac{1}{(2\pi)^2}\int d^{2}k d^{2}y
e^{i\overrightarrow{k}\cdot
(\overrightarrow{x}-\overrightarrow{y})}f(\overrightarrow{y}),
\label{2}
\end{equation}
by the corresponding operator $\hat{\overrightarrow{x}}$ :
\begin{equation}
W[f(x_{i})] = \frac{1}{(2\pi)^2}\int d^{2}k d^{2} y
e^{i\overrightarrow{k} \cdot (\hat{\overrightarrow {x}} -
\overrightarrow{y})}f(\overrightarrow{y}).\label {3}
\end{equation}
For example, if $f(\overrightarrow {x}) = x_1 x_2$, then one can
easily show that
\begin{equation}
W[x_1 x_2] = \frac{1}{2} (\hat x_1 \hat x_2 + \hat x_{2} \hat
x_{1}). \nonumber
\end{equation}
One can then show that for any pair of such functions $f(x_{i})$
and $g(x_{i})$ the composition rule between them should be
modified appropriately to the Moyal star product (*), so that
\begin{equation}
W[f(x_{i})]W[g(x_{i})] = W[(f*g)(x_{i})]\label{4}
\end{equation}
where
\begin{equation}
(f*g)(x_{i}) = e^{\frac{i}{2}\theta \epsilon_{ij}
\partial_{i}^{x}\partial_{j}^{y}}
f(x_{i})g(y_{i})|_{y=x}.\label{5}
\end{equation}
For a more general formulation see \cite{kupr}. It is quite easy to verify the Moyal brackets defined through this
star product, between the c-numbered valued coordinates now
becomes isomorphic to the commutators of their corresponding
operators (\ref{1})
\begin{equation}
[x_{1},x_{2}]_{*} = x_{1}*x_{2} - x_{2}*x_{1} = i\theta. \label{6}
\end{equation}
Although not commutative, the Moyal star product is associative.

   This 2D noncommutative problem arises naturally in the Landau problem, where a charged particle (say an electron) moving
in $xy$ plane and subjected to a transverse magnetic field 'B'
along the z-direction. In this problem, which plays a central role
in QHE the commuting $\hat x_{1}$ and $\hat x_{2}$ coordinates
fails to commute, when projected to the lowest Landau level
\cite{szabo,2}
\begin{equation}
[P\hat x_{1}P, P\hat x_{2}P] = \frac{1}{iB} \label{7}
\end{equation}
so that the noncommutativity is given by $(\frac{1}{B})$. This
noncommutativity implies that the guiding center coordinates in
the Lowest Landau Level  satisfy the usual uncertainty relations
\begin{equation}
\Delta x_{1} \Delta x_{2} \ge \frac{1}{B}
\end{equation}
This uncertainty relations between the projected coordinates (here we have intentionally
suppressed the projection operator $P$) indicates that the minimal
area occupied by any particle is $\approx \frac{1}{B}$ in Lowest Landau Level  and,
thus for fermions there exists an upper bound to the number of
particles that can be accommodated in unit area. On the other
hand, interparticle interactions can renormalise the
noncommutative parameter away from $(\frac{1}{B})$ and thus can
have affect on the filling fraction in a Quantum Hall system.
Indeed, it has been demonstrated that Jain fraction for Fractional Quantum Hall Effect can
be obtained in a heuristic treatment, where the electrons are
attached to appropriate magnetic flux tubes  \cite{scholtz}. Furthermore, there are certain situations where interactions can be traded with noncommutativity within a certain approximation, as can be seen by constructing a dual families of noncommutative quantum systems \cite{newref}.
\subsection{Problem Regarding Poincare/Galilean Symmetry and its Restoration}
The expression of Moyal star-product given above (\ref{5}) can  be
generalised to $3+1$ dimensional spacetime as
\begin{equation}
(f*g)(x)=e^{\frac{i}{2}\theta^{\mu\nu}\partial_{\mu}^{x}\partial_{\nu}^{y}}
f(x)g(y)|_{y=x} \label{8}
\end{equation}
However, the introduction of the length scale through the NC
matrix $\Theta = \{\theta^{\mu\nu}\}$ in more than two-dimension
violates Poincare or more precisely, the symmetry under
homogeneous Lorentz transformation. This can be understood easily
by considering the transformation property of scalar field
$\phi(x)$ under a homogeneous Lorentz transformation
\begin{eqnarray}
x^{\mu} &\rightarrow& x'^{\mu} = \Lambda^{\mu}_{\quad \nu} x^{\nu} \\
\phi &\rightarrow&\phi^{\Lambda}(x) =
\phi(\Lambda^{-1}x)\label{9'}
\end{eqnarray}
One can then easily see that for a pair of arbitrary scalar fields
$\phi_{1}(x)$ and $\phi_{2}(x)$, the automorphism under the
homogeneous Lorentz group does not hold
\begin{equation}
(\phi_{1}^{\Lambda} * \phi_{2}^{\Lambda})(x) \ne
(\phi_{1}*\phi_{2})^{\Lambda}(x) \label{10'}
\end{equation}
Infact the easiest way to understand (\ref{10'}) is through (\ref{02}).
(Note that the translational symmetry is not affected and
like-wise in $D=2$ case, considered above, the $SO(2)$ symmetry is
also not violated, as $\theta_{ij}$ behaves as an $SO(2)$
scalar.). This symmetry can, however, be restored by using an
appropriate Drinfeld twist in a Hopf algebraic framework. To
illustrate the essential ideas involved, let us consider the case
of $SO(3)$ symmetry itself in $R^{3}$, where we take $\theta^{0i}
=0$. Being a subgroup also of the Galilean group, this demonstrates
how the Galilean symmetry can also be restored in the framework.
In fact, even in the presence of a nonvanishing $\theta^{0i}$ the
broken symmetry under Galilean boost is taken care of rather
trivially \cite{4}. To begin with, one interprets the NC $*$
-product introduced in (\ref{8}) as to correspond NC algebra
$\mathcal{A}_{\theta}$, obtained by deforming the commutative
algebra $\mathcal{A}_{o}$, where fields compose  through
point-wise multiplication, i.e effectively with $\theta =0$.

   More formally, the multiplication map '$m_{\theta}$' in $\mathcal{A}_{\theta}$ is obtained from '$m_{o}$' in $\mathcal{A}_{o}$
\begin{eqnarray}
m_{o}: \mathcal{A}_{o}\otimes \mathcal{A}_{o} &\rightarrow& \mathcal{A}_{o} \label{9}\\
m_{o}(f(x)\otimes g(x))  &=& f(x)g(x) \nonumber
\end{eqnarray}
by inserting a twist operator $\mathcal{F}_{\theta}\equiv
e^{-\frac{i}{2}\theta^{ij}P_{i}\otimes P_{j}} \in
\mathcal{U}(ISO(3))\otimes\mathcal{U}(ISO(3))$ as
\begin{eqnarray}
m_{\theta}: \mathcal{A}_{\theta}\otimes \mathcal{A}_{\theta} &\rightarrow& \mathcal{A}_{\theta} \label{10}\\
m_{\theta}(f(x)\otimes g(x))  &=&
m_{0}(\mathcal{F}_{\theta}(f(x)\otimes g(x)))\nonumber
\end{eqnarray}
The operators $P_i$ used above are the generators of the translational algebra $T_3$. Here $ ISO(3)= SO(3) \ltimes T_{3}$ is  a semi-direct product
of $SO(3)$ and the translational algebra $T_{3}$ in ${\mathbb{R}}^{3}$
and $\mathcal{U}(ISO(3))$ is the corresponding universal
enveloping algebra, which is a Hopf algebra. This suggest that one
has to go beyond the usual Lie-algebraic framework to a (deformed)
Hopf algebraic one to capture the relevant symmetries. Indeed, it
can be easily shown that the usual co-product of the deformed Hopf
algebra should also be deformed as \cite{7,wess} (for a review see \cite{5}).
\begin{equation}
\Delta_{o}(\overrightarrow{J})= \overrightarrow{J} \otimes
\mathbb{I} + \mathbb{I}\otimes \overrightarrow{J}\rightarrow
\Delta_{\theta}(\overrightarrow{J}) = \mathcal{F}_{\theta}^{-1}
\Delta_{o}(\overrightarrow{J})\mathcal{F}_{\theta} \label{11}
\end{equation}
in order to be compatible with the deformed product '$m_{\theta}$'
(\ref{8}):
\begin{equation}
m_{\theta}[\Delta_{\theta}(\overrightarrow{J})(f\otimes g)] =
\overrightarrow{J} \triangleright m_{\theta}(f\otimes g).
\label{12}
\end{equation}
Here '$\overrightarrow{J}$' is taken to be the $SO(3)$ generators.
Note that the co-product of the translational generators
$\overrightarrow{P} \in T_{3}$ does not undergo any deformation by
this abelian twist:
\begin{equation}
\Delta_{o}(\overrightarrow{P}) \rightarrow
\Delta_{\theta}(\overrightarrow{P}) =
\mathcal{F}_{\theta}^{-1}\Delta_{o}(\overrightarrow{P})\mathcal{F}_{\theta}
= \Delta_{o}(\overrightarrow{P}) \label{13}
\end{equation}
Besides, the other maps the antipode and the co-unit undergoes no
deformation. An immediate consequence of this deformation is that
two particle exchange map $\tau_{o}(\phi\otimes \psi)= \psi\otimes
\phi$ and hence the corresponding projection operator
 $P_{o}=\frac{1}{2}(\mathbb{I}\pm \tau_{o})$ into the symmetric/antisymmetric subspaces,
 describing bosons/fermions, also get deformed
in a similar manner \cite{6}
\begin{eqnarray}
\tau_{o} &\rightarrow& \tau_{\theta} = \mathcal{F}_{\theta}^{-1}\tau_{o}\mathcal{F}_{\theta}\\
P_{o} &\rightarrow& P_{\theta} =
\mathcal{F}_{\theta}^{-1}P_{o}\mathcal{F}_{\theta}\label{14}
\end{eqnarray}
In the process one introduces the, so-called, twisted
boson/fermions through the deformed exchange operator
$\tau_{\theta}$, ensuring super-selection principle, as
\begin{equation}
[\Delta_{\theta},\tau_{\theta}] = [\Delta_{\theta},P_{\theta}] = 0
\label{15}
\end{equation}
Furthermore, it has also been shown that $\theta_{ij}$ indeed
transforms as a scalar under the action of SO(3), when it is
implemented through the above twisted co-product \cite{7}
\begin{equation}
\overrightarrow{J}_{\theta} \triangleright \theta^{ij} = 0
\label{16}
\end{equation}
where the subscript '$\theta$' indicate that one has to consider
the action of $\overrightarrow{J}$, through a twisted co-product.
Finally, let us demonstrate through a heuristic argument, how the
algebra involving the corresponding creation/annihilation
operators also get deformed \cite{4,8} which was observed for the
first time in \cite{6} .

   To begin with, let us apply the twisted projection operator $P_{\theta}$ on the tensor product of two momentum
   eigenstates
 $(|k\rangle \otimes |l\rangle)$ to write it in the usual symmetric/antisymmetric form up to a phase factor as
\begin{equation}
P_{\theta} (|k\rangle \otimes |l\rangle) =
\frac{1}{2}(|k\rangle\otimes |l\rangle \pm
\mathcal{F}^{-2}_{\theta}|l\rangle\otimes |k\rangle) = \frac{1}{2}e^{ik
\wedge l}(|k,l\rangle\rangle \pm |l,k\rangle\rangle) \label{17}
\end{equation}
where
\begin{equation}
|k,l\rangle\rangle \equiv e^{-ik\wedge l}|k\rangle \otimes
|l\rangle
\end{equation}
and
\begin{equation}
k\wedge l = \frac{1}{2}\theta^{ij}k_{i}l_{j} \label{18}
\end{equation}
Now identifying $a_{k}^{\dagger}a_{l}^{\dagger}|0\rangle =
P_{\theta}(|k\rangle\otimes |l\rangle)$, it easily follows that
\begin{equation}
a_{k}^{\dagger}a_{l}^{\dagger} = \pm e^{2ik\wedge
l}a_{l}^{\dagger}a_{k}^{\dagger}. \label{19}
\end{equation}
The phase of the other commutation relation
\begin{equation}
a_{k}^{\dagger}a_{l} = \pm
(a_{l}a_{k}^{\dagger}-(2\pi)^{3}\delta^{3}(\overrightarrow {k}
-\overrightarrow{l}))e^{-2ik\wedge l} \label{20}
\end{equation}
can be understood easily from the fact that the annihilation
operator $a_{l}$ is associated with momentum $(-l)$, in contrast
to the operator $a_{l}^{\dagger}$, for which the associated
momentum is ${+l}$. These operators, which satisfy a
(anti)commutation relation turns out to be related to the
undeformed $(\theta =0$ ones through a $U(1)$ transformation as
$a_k=a_k(\theta =0 )exp(\frac{i}{2}k\wedge P )$ where $P$ is the
total momentum. For an interpretation through deformed Heisenberg
algebra see \cite{topan}.

  These twisted (anti) commutation relations can have some of the drastic consequences. For example,
  it can violate Pauli's exclusion
principle \cite{4}. To illustrate this briefly, consider the
density matrix of a canonical ensemble, comprising a pair of free
identical twisted fermions/bosons: 
\begin{equation}
\rho = e^{-\beta H} ; H = \frac{1}{2m}(\overrightarrow{ P}^{2}
\otimes \mathbb{I}+ \mathbb{I}\otimes
\overrightarrow{P}^{2}).\label{21}
\end{equation}
The 2-particle correlation function in the "thermodynamical" limit
is then obtained as
\begin{equation}
C(r=|\overrightarrow{r_{1}} - \overrightarrow{r_{2}}|) =
\frac{1}{Z} ((\langle\overrightarrow{r_{1}}|\otimes
\langle\overrightarrow{r_{2}}|)P_{\theta}\rho P_{\theta})
(|\overrightarrow{r_{1}}\rangle\otimes
|\overrightarrow{r_{2}}\rangle) \label{22}
\end{equation}
which represents the probability density of a particle to be
detected around $\overrightarrow{r_{1}}$, given that the other is
at $\overrightarrow{r_{2}}$.
 This in turn relates to the effective (statistical) potential $V_{eff}(r)$ as
\begin{equation}
C(r) = e^{-\beta V_{eff}(r)}
\end{equation}
(see \cite{4} and references there in).

  On explicit computation, one easily finds that for a $2D$ system a large area $A$, this is given by
\begin{equation}
C(r)= \frac{1}{A^{2}}\left[1\pm
\frac{1}{1+\frac{\theta^{2}}{\lambda^{4}}}e^{\frac{-2\pi
r^{2}}{[\lambda^{2}(1+\frac{\theta^{2}}{\lambda^{4}})]}}\right]\label{23}
\end{equation}
where $\lambda = \sqrt{\frac{2\pi\beta}{m}}$ is the thermal
wavelength and $\beta=\frac{1}{kT}$ and $\pm $ sign referes to twisted bosonic and fermionic cases respectively.

The profile of $V_{\mathrm{eff}}(r)$  for twisted fermions (TF)
and the conventional fermions ($\theta = 0$) shows that the
$V_{\mathrm{eff}}$ saturates to a finite value (soft core
potential) for coincident points of a part of twisted fermions,
unlike the usual case, where it diverges (hard core). This
demonstrates that two fermions can indeed sit on the top of each
other, in principle, violating Pauli's exclusion principle,
although it requires an enormous energy to do so. For other
effects, see \cite{5}.

We would, however, like to mention here that there were some controversies recently in the literature on this issue of twisted statistics. Particularly in \cite{fiore} it was pointed out that one should implement braided twisted symmetry, rather than the usual twisted symmetry. In this scheme one is entitled to define $*$-product between fields even at distinct spacetime points, which virtually washes out any noncommutative effect. In particular, the twisted symmetric fermions or bosons  do not occur here. In fact, it has been shown in \cite{8}  that the energy shift of noncommutative origin does not appear in the system of degenerate electron gas, if braided twisted symmetry is implemented, whereas it occurs in the case of usual twisted symmetry. But this was again criticised in a recent work \cite{pi} (see also \cite{balanewref}).

\subsection{Operatorial Approach}

So far, we have been dealing with Moyal star product formalism in
dealing with the various quantum mechanical/field theoretical
systems. On the other hand, recently a formalism is being
developed \cite{9} in the context of noncommutative quantum mechanics, where one confronts the operatorial nature of
the coordinate variables ``head on," rather than `demoting' them
to ordinary c-number variables and use Moyal star product{\footnote{It should be recalled that there is an essential difference between quantum mechanics and quantum field theory, as far as the status of the position coordinates are concerned: Even in the ordinary commutative $(\theta =0)$ theories, the role of position coordinates $\hat x_i$ is that of an hermitian observable, whereas in quantum field theory these are mere labels of the continuous degrees of freedom and therefore they do not belong to the configuration space and do not satisfy any Heisenberg algebra. Interestingly, this difference carries over (as expected) to the noncommutative counterparts. Although the status of $x_i$'s is upgraded to an operator $\hat x_i$ in noncommutative theories it is definitely not  valued in the same vector space as that of the field operators in noncommutative quantum field theories.}}.  So far
this has been sorted out for the spatial dimension $D=2$ only and
work is in progress to extend the formalism to $D=3$ and beyond.
To cite  examples two major accomplishments of the framework is that it has
been possible to find the exact energy eigenvalues of a simple
quantum mechanical problem of a particle confined in an infinite
spherical potential well. It will indeed be very difficult, if not
impossible, to obtain the same in a Moyal product formalism. Besided in \cite{sunan} the authors have provided a path-integral representation of the transition amplitude, using coherent state approach, and obtained the form of a non-local action of a particle moving in noncommutative plane subjected to arbitrary potential.

In the following we outline briefly the essential ideas involved.
To begin with, one observes that the structure of the coordinate
algebra $[\hat{x}, \hat{y}]=i \theta$ in $D=2$ case, is isomorphic
to the phase space Heisenberg algebra $[\hat{x}, \hat{p}]=i \hbar$
of $1-D$ harmonic oscillator so that here $\hat{y}$ plays the role
of $\hat{p}$, while $\theta$ plays the role of $\hbar$.
Consequently the classical configuration space $\mathcal{H}_c$ in the NC
case is just isomorphic to the bosonic Fock space of the harmonic
oscillator. One thus defines the classical configuration space as
\begin{eqnarray}
\label{26} \mathcal{H}_c=\mathrm{Span}\lbrace\vert
n\rangle\rbrace_{n=0}^\infty
\end{eqnarray}
where the span is taken over the field of complex numbers and
\begin{eqnarray}
\label{27} \vert n\rangle = \frac{1}{\sqrt{n!}}
\left(b^\dagger\right)^n \vert 0\rangle ;\qquad\qquad
b=\frac{1}{\sqrt{2\theta}}\left( \hat{x}+i\hat{y}\right)
\end{eqnarray}
and $\vert 0 \rangle$ satisfies $b\vert 0 \rangle=0$ by
definition.

The quantum Hilbert space, (the set of all quantum states) is
identified with
\begin{eqnarray}
\label{28} \mathcal{H}_q = \left\lbrace  \psi
\left(\hat{x},\hat{y}
\right):\textrm{tr}_c\Big(\psi^\dagger\left(\hat{x},\hat{y}
\right)\psi\left(\hat{x},\hat{y} \right)\Big)<\infty\right\rbrace
\end{eqnarray}
In other words, the Hilbert space is the trace class enveloping
algebra of the classical configuration space $\mathcal{H}_c$ Fock
algebra $\left(b,b^\dagger\right)$. As these operators are
necessarily bounded, this is again a Hilbert space (recall that
the set of all bound operators in a Hilbert space is again a
Hilbert space). We denote the states in $\mathcal{H}_c$ by $\vert
\centerdot \rangle$, whereas a state in $\mathcal{H}_q$ will be
denoted by $\vert \centerdot )$. The corresponding inner product
is $\left( \psi\vert\phi\right)=\left( \psi ,
\phi\right)=\textrm{tr}_c\left( \psi^\dagger\phi\right)$, which
also serves to define bra states as elements of the dual space
(linear functionals). Note that the trace  is performed over the
classical configuration space, denoted by the subscript `$c$'.

The next step is to construct a representation\cite{cite1,cite2} of the NC
Heisenberg algebra
\begin{eqnarray}
\label{29} [\hat{x}_i^q,\hat{x}_j^q]=i \theta
\epsilon_{ij}\,;\qquad [\hat{x}_i,\hat{p}_j^q]=i \hbar\
\delta_{ij},;\qquad [\hat{p}_i^q,\hat{p}_j^q]=0 \qquad\quad
(i,j=1,2)
\end{eqnarray}
on $\mathcal{H}_q$. This is done simply by defining the action of
these operators as follows:
\begin{eqnarray}
\label{30}
\hat{x}_i^q \psi \left(\hat{x},\hat{y} \right)=\hat{x}_i\psi \left(\hat{x},\hat{y} \right)\nonumber\\
\hat{p}_i^q \psi \left(\hat{x},\hat{y}
\right)=\frac{\hbar}{\theta} \epsilon_{ij}\left[\hat{x}_j,\psi
\left(\hat{x},\hat{y} \right)\right]
\end{eqnarray}
where $\psi \left(\hat{x},\hat{y} \right)$ is an arbitrary
operator in the $\mathcal{H}_q$. Note that the momenta act
as inner automorphisms with respect to $\hat{x}_i$ and $\hat{x}_i$'s are taken to act through left
multiplication. It can now be trivially verified, by using the NC
Heisenberg algebra (\ref{29}), the definition of the inner product
$\left( \psi\vert\phi\right)$ and the Jacobi identity, that it, in
fact, furnishes a unitary representation. For quantum mechanical
interpretation of this formalism, see \cite{10}.

Using this formalism, the complete spectrum of a particle,
confined in a NC spherical well in $D=2$ could be obtained
\cite{9}. The bound and scattering states for finite potential
barrier could also be obtained unambiguously. It was also shown
that the time-reversal symmetry is broken by noncommutativity,
which can only be restored in the commutative or thermodynamic
limit. Based on this, the thermodynamics of an ideal Fermi gas in
this infinite spherical well has also been studied \cite{11},
which exhibits some of the remarkable behaviors implied by the
excluded area resulting from the noncommutativity. In particular,
there are extremal macroscopic states, characterized by area,
number of particles and angular momentum, that correspond to a
single microscopic state and thus have vanishing entropy.
Furthermore, for comparable system size and excluded area, the
thermodynamical quantities, such as entropy, exhibit non-extensive
features.

\section{Doubly Special Relativity and Noncommutativity }
The idea of Doubly (or Deformed) Special Relativity (DSR),
formulated by Amelino-Camelia \cite{am} was a culmination of
several apparently disconnected issues. The combined wisdom of
theorists demanded a radical departure from conventional physics
in the regime of Planck scale \cite{planck}. It appeared
\cite{mdis} that one needs to modify the (Einsteinean ) energy
momentum dispersion law, a possible form being $E^2=c^2\vec p^2
+c^4m^2 +\eta L^n_Pc^2\vec p^2E^n +...$, where $\eta$ is a
numerical factor and $L_P$ is a fundamental length scale, which
can be Planck length. The modification has to be such that at low
energy the standard relation $E^2=c^2\vec p^2 +c^4m^2$, compatible
with Special Theory (SR),  is recovered.

But this leads to the first clash with SR because (Planck) length
or energy are not observer {\it{independent}} quantities. In SR
$c$ is the only observer independent scale. However we do not want
to discard the Relativity Principle: The laws of physics take the
same form in all inertial frames. The only way to achieve both the
above is to generalize the SR (coordinate and momentum)
transformation rules in such a way that {\it{two}} observer
independent scales $c$ and $L_P$ (instead of the single one $c$ in
SR) can be accommodated - hence Doubly (or Deformed) Special
Relativity (DSR) \cite{am}. The situation is similar to the
transition from no scale and linear  Galilean Relativity to one
scale ($c$) SR where the velocity addition theorem becomes
non-linear. In a similar vein the transition from SR to DSR yields
two scales ($c$ and $L_P$) at the cost of ushering another level
of non-linearity, that in the energy momentum transformation
rules. In fact we will see that the phase space variable
transformation rules and invariants under DSR laws get completely
entangled \cite{mag,gho2}.

However there is another very important feature of DSR: it is
intimately connected to a Non-Commutative spacetime, the
$\kappa$-Minkowski spacetime \cite{am,dsr2}. This is indeed nice
because a theory claimed to be valid at Planck length has to have
an inherent noncommutativity because the Quantum Gravity models
\cite{planck} predict a foam like discrete spacetime below Planck
length. Furthermore the coexistence of black hole physics and
quantum mechanics also demands a discrete spacetime \cite{dfr}.

Since our focus is on the NC aspect of DSR we will not discuss the
DSR itself anymore (for which there are reviews \cite{rev}). We
will concentrate more on the NC aspect of DSR. Again there are two
popular ways to introduce noncommutativity in DSR scenario: (I)
The $\kappa $-Poincare Hopf algebra approach \cite{kpon} and the
$\kappa $-Minkowski spacetime approach \cite{dsr2}. In the former
the Poincare Lie algebra framework of SR is extended to a Hopf
algebra. In the latter one tries to keep the SR Poincare algebra
intact by changing the Poincare generators and transformation laws
appropriately. One advantage of the latter \cite{gho2} is that the
Lorentz group theoretic classifications of the quantum fields in
DSR will possibly remain unchanged. In the present article we will
follow the second alternative.
\subsection{DSR Phase Space}
 In the sense of classical Poisson
Brackets, the NC $\kappa$-Minkowski spacetime is defined as
\cite{am,dsr2,gho2},
\begin{equation}
\{x^i,x^0\}=\frac{x^i}{\kappa} ~;~~ \{x^i,x^j\}=0. \label{ok}
\end{equation}
 The form of of modified
dispersion relation, we will consider was proposed by
 Magueijo and Smolin  \cite{mag} (MS)
\begin{equation}
p^2= m^2[1-\frac{E}{\kappa} ]^2 =  m^2[1-\frac{(\eta p)}{\kappa}
]^2 \label{0m}
\end{equation}
with $E$ being the particle energy and $(\eta p)=\eta ^\mu p_\mu~,
\eta _0=1,\eta _i=0$. Here the metric is diagonal with components $g^{00}=-g^{ii}=1$.

It should be mentioned that, even if one imposes the restrictions
that Jacobi identities have to be maintained and that the
structure should reduce to canonical algebra for $\kappa
\rightarrow \infty $, the full $\kappa$-NC phase space algebra  is
not uniquely determined. There are distinct (and possibly
inequivalent) representations that are connected by non-linear
transformations \cite{kow}. This is possible because DSR formalism can be interpreted as a nonlinear realization of the Lorentz group. It is interesting to point out that for a particular combination of variables\cite{kow} the DSR algebra becomes isomorphic to the Snyder algebra \cite{sn}. However the  sector of the algebra
common to all forms of $\kappa$-Minkowski spacetime is
\begin{equation}
\{x^i,x^0\}=\frac{x^i}{\kappa} ~;~~
\{x^i,x^j\}=0~;~~\{x^i,p^j\}=-g^{ij}~;~\{p^\mu,p^\nu\}=0.
\label{ko}
\end{equation}
The particular $\kappa$-NC phase space that we will use here was
first studied and further developed in \cite{granik} (in a
restricted set up of $1+1$-dimensional toy model). In fact this
phase space can be extracted from very general deformations
considered by Lukierski et.al. \cite{dsr2}. Rest of the phase
space algebra is given below,
\begin{equation}
 \{x^0,p^i\}=p^i/\kappa ~;~\{x^i,p^0\}=0~;~\{x^0,p^0\}=-1+p^0/\kappa.
\label{002}
\end{equation}
The above  is rewritten in a covariant form,
$$
\{x_\mu ,x_\nu \}=\frac{1}{\kappa}(x_\mu \eta_{\nu}-x_\nu
\eta_{\mu }),$$
\begin{equation}
\{x_{\mu},p_{\nu}\}=-g_{\mu\nu}+\frac{1}{\kappa}\eta_{\mu}p_{\nu},~~\{p_{\mu},p_{\nu}\}=0.
 \label{03}
\end{equation}

 We wish to construct the finite Lorentz transformation (LT) consistent with this NC space.
 These were first constructed in \cite{kimb}. We will instead follow
 another more systematic route that was exploited in \cite{bru,gho2}. We
 need the rotation generators to generate infinitesimal variations. The angular momentum is defined in the normal way as,
\begin{equation}
J_{\mu\nu }=x_\mu p_\nu -x_\nu p_\mu . \label{j}
\end{equation}
This is motivated by the fact that spatial sector of $\kappa$-NC
algebra in (\ref{03}) remains unaffected. Furthermore,  using
(\ref{03}) one can check that the Lorentz algebra is intact,
\begin{equation}
\{J^{\mu\nu },J^{\alpha\beta }\}=g^{\mu\beta }J^{\nu\alpha
}+g^{\mu\alpha }J^{\beta \nu}+g^{\nu\beta }J^{\alpha\mu
}+g^{\nu\alpha }J^{\mu\beta }. \label{51}
\end{equation}
 However,
Lorentz transformations of $x_\mu$ and $p_\mu $ are indeed
affected,
\begin{equation}
\{J^{\mu\nu},x^\rho \}=g^{\nu\rho}x^\mu-g^{\mu\rho}x^\nu
+\frac{1}{\kappa } (p^\mu\eta^\nu -p^\nu \eta^\mu )x^\rho ~;~
\{J^{\mu\nu},p^\rho \}=g^{\nu\rho}p^\mu -g^{\mu\rho}p^\nu
-\frac{1}{\kappa } (p^\mu\eta^\nu -p^\nu \eta^\mu )p^\rho .
\label{52}
\end{equation}
Notice that the extra terms appear  for $J^{0i}$ and not for
$J^{ij}$ so that only boost transformations are changed.

From now on we will use the $(x,y,z,t)$ notation (instead of the
covariant one), which is more suitable for comparison with
existing results. We define the infinitesimal transformation of a
generic variable $O$ by,
\begin{equation}
\delta O=\{\frac{1}{2}\omega_{\mu\nu}J^{\mu\nu},O\}, \label{l}
\end{equation}
and only the parameter $\omega_{0x}=\delta u$ is non-vanishing.

Let us start with the energy-momentum vector $(E,p_x,p_y,p_z)$.
The above considerations yield the following differential
equations \cite{gho2},
\begin{equation}
\frac{dE}{du}=-p_x+\frac{Ep_{x}}{\kappa};~~\frac{dp_x}{du}=-E+\frac{p^{2}_{x}}{\kappa};~~\frac{dp_y}{du}=\frac{p_{y}p_{x}}{\kappa};~~\frac{dp_z}{du}=\frac{p_{z}p_{x}}{\kappa}.
\label{l1}
\end{equation}
The details can be obtained from \cite{gho2}. The final result is
the $\kappa $-LT rules,
\begin{equation}
E'=\frac{\gamma (E-vp_x)}{\alpha};~p'_x=\frac{\gamma
(p_x-vE)}{\alpha};~p'_y=\frac{p_y}{\alpha};~p'_z=\frac{p_z}{\alpha}
\label{l10}
\end{equation}
where $\gamma = \frac{1}{\sqrt{1-v^2}},~\alpha
=1+\frac{1}{\kappa}\{((\gamma -1)E-v\gamma p_x\})$. Notice that,
unlike SR LTs, in $\kappa $-LT the components transverse to the
velocity $v$ are also affected and the $\kappa $-effect  appears
as the factor $\alpha $ and for $\kappa =\infty ,~ \alpha =1$ so
that SR LTs are recovered.

Before proceeding to derive the $\kappa$-LT for the coordinates
$x_\mu $, let us first find out the new dispersion law that is
$\kappa$-LT invariant. Scanning the following infinitesimal
transformation rules,
\begin{equation}
\{\frac{1}{2}J_{\mu\nu},p^2\}=\frac{p^2}{\kappa}(\eta _\mu p_\nu
-\eta _\nu p_\mu );~~\{\frac{1}{2}J_{\mu\nu},(\eta
p)\}=-(1-\frac{(\eta p)}{\kappa})(\eta _\mu p_\nu -\eta _\nu p_\mu
);~~ \label{l11}
\end{equation}
we find the following combination to be invariant:
\begin{equation}
\{\frac{1}{2}J_{\mu\nu},\frac{p^2}{(1-\frac{(\eta
p)}{\kappa})^2}\}=0. \label{l12}
\end{equation}
The finite $\kappa$-LTs also yields
\begin{equation}
(p^2-m^2(1-\frac{(\eta p)}{\kappa})^2)'=\frac{1}{\alpha
^2}(p^2-m^2(1-\frac{(\eta p)}{\kappa})^2), \label{l13}
\end{equation}
confirming that the new $\kappa$-LT invariant dispersion law is
\begin{equation}
p^2=m^2(1-\frac{(\eta p)}{\kappa})^2. \label{l14}
\end{equation}
This is the MS dispersion law \cite{mag}  (\ref{0m}).

In an identical fashion, putting the coordinates for $O$ in
(\ref{1}) we compute the $\kappa $-LTs for the coordinates
\cite{kimb,gho2},
\begin{equation}
t'=\alpha \gamma (t-vx);~~x'=\alpha \gamma (x-vt),~~y'=\alpha
y,~~z'=\alpha z. \label{l21}
\end{equation}
Once again we notice the similar features as in the momentum
transformation laws.

As in the dispersion relation, we look for an invariant quantity
that will generalize the conventional distance and we find that
under the $\kappa$-LT (\ref{l21}),
\begin{equation}
(x^2(1-\frac{(\eta p)}{\kappa})^2)'=x^2(1-\frac{(\eta
p)}{\kappa})^2. \label{l22}
\end{equation}
Hence the invariant length is generalized to
\begin{equation}
s^2=x^2(1-\frac{(\eta p)}{\kappa})^2. \label{k1}
\end{equation}
This is one of the important results of \cite{gho2} that can have
connections with Finsler geometry \cite{fins}. Its
$1+1$-dimensional analogue was suggested by Mignemi in \cite{granik}. \\
\subsection{Canonical Variables}
Now we will introduce a new set of phase space variables which
obey canonical Poisson brackets are transform in the conventional
way under SR Lorentz transformation. Somewhat similar
considerations in parts have appeared before in \cite{granik} but
exhaustive study of the full canonical phase space was given in
\cite{gho2}. Indeed,  these variables are composites of phase
space coordinates will have to suitable ordered upon quantization.
But, in the classical framework they will prove to be very
convenient and they drastically simplify the computations while
analyzing phenomenological consequences of the modified Lorentz
transformations. We will return to the quantum case at the end.

The two invariant quantities that we derived in
(\ref{0m},\ref{k1}) suggest the forms of these canonical avatars:
\begin{equation}
X_\mu \equiv x_\mu (1-\frac{(\eta p)}{\kappa})=x_\mu
(1-\frac{E}{\kappa});~~P_\mu \equiv \frac{p_\mu}{(1-\frac{(\eta
p)}{\kappa})}=\frac{p_\mu}{(1-\frac{E}{\kappa})}. \label{c1}
\end{equation}
We remind  that the variables on the right hand side obey
$\kappa$-LT laws. Using the NC algebra (\ref{03}) it is easy to
check the following:
\begin{equation}
\{X_\mu ,P_\nu \}=-g_{\mu\nu};~~\{X_\mu ,X_\nu \}=\{P_\mu ,P_\nu
\}=0. \label{c2}
\end{equation}
Hence the $X,P$ phase space is canonical. The above relations in
(\ref{c1}) are invertible,
\begin{equation}
x_\mu = X_\mu (1+\frac{(\eta P)}{\kappa})=X_\mu
(1+\frac{P_0}{\kappa});~~p_\mu = \frac{P_\mu}{(1+\frac{(\eta
P)}{\kappa})}=\frac{P_\mu}{(1+\frac{P_0}{\kappa})}. \label{c3}
\end{equation}
Next we consider Lorentz transformations of the canonical
variables and find, for example,
$$T'=t'(1-\frac{E}{\kappa})'=\gamma \bar \alpha (t-vx)[1-\frac{\gamma}{\kappa \bar \alpha}(E-vp_x)]$$
\begin{equation}
=\gamma [t(1-\frac{E}{\kappa})-vx(1-\frac{E}{\kappa})]=\gamma
(T-vX),$$
$$P'_0=\frac{E'}{(1-\frac{E}{\kappa})'}=\gamma
(P_0-vP_x). \label{c4}
\end{equation}
where $\bar \alpha =\alpha (-v)$ and we have used the identity
$(\bar \alpha )^{-1}=\alpha '$. One can easily check that all the
canonical variables $X_\mu,P_\mu $ obey SR LTs.

Before putting to use this canonical variables we should
add a cautionary remark. This mapping has led to conjectures
\cite{garcia} that in DSR the NC phase space formulation is
redundant. However this mapping is classical and the quantization
of this mapping will require a proper representation of the NC
operators and this is an open problem. Some aspects have appeared
in \cite{wess,mel} in the context of $\kappa$-Minkowski quantum
field theory. Furthermore the NC quantum theories are
qualitatively different with an inherent non-locality that
obviously can not be captured in the canonical set up which,
however, provides a very convenient framework to construct the
quantum theory. An explicit example is given \cite{goss} where we
construct  the DSR generalization of Dirac fermions in a very
simple way as compared to the original derivation \cite{am1}.\\

\subsection{Deformed Symmetry Generators}
  Let us show how to exploit the classical phase space
in constructing deformed symmetry generators \cite{gho2} (for the
Snyder algebra see \cite{rb}). In the conventional case, the phase
space algebraic structure of the point particle is invariant under
the following symmetry transformations: translation, Lorentz
rotation, dilation and special conformal transformation. On the
other hand, the particle dispersion relation $P^2-m^2=0$ enjoys
invariance under translation and Lorentz rotation, and the mass
term $m$ breaks the symmetry under dilation and special conformal
transformation. Finally, the symmetry generators satisfy a closed
algebra among themselves.

In the $\kappa$-particle model our aim is to construct the
generators in the $\kappa$-NC space that preserve invariances of
both the $\kappa$-NC phase space algebra (\ref{03})  and the
structure of the algebra among generators (see below in
(\ref{0s})). Then we will check how the $\kappa$-modified
dispersion relation (MS relation (\ref{0m}) in the present case)
is affected. Once again the canonical $(X_\mu ,P_\mu )$ variables
will do the trick. The idea is to first write down the generators
in terms of $(X_\mu ,P_\mu )$ degrees of freedom using the
conventional form of the generators, ({\it{i.e.}} that of normal
particle in normal phase space). They will obviously satisfy the
standard closed algebra among generators:
$$
\{J^{\mu\nu },J^{\alpha\beta }\}=g^{\mu\beta }J^{\nu\alpha
}+g^{\mu\alpha }J^{\beta \nu}+g^{\nu\beta }J^{\alpha\mu
}+g^{\nu\alpha }J^{\mu\beta }~;~~\{J^{\mu\nu },T^\sigma
\}=g^{\nu\sigma}T^{\mu}-g^{\mu\sigma}T^{\nu}~; $$
$$ \{J^{\mu\nu },D \}=0~;~~ \{J^{\mu\nu },K^\sigma \}=2D(g^{\nu\sigma}X^{\mu}-g^{\mu\sigma}X^{\nu})-X^2(g^{\nu\sigma}T^{\mu}-g^{\mu\sigma}T^{\nu})~;$$
$$\{T^\mu ,T^\nu \}=0~;~~\{T^\mu ,D\}=T^\mu ~;~~\{T^\mu ,K^\nu \}=2Dg^{\mu\nu}-2J^{\mu\nu}~;$$
\begin{equation}
\{D ,D \}=0~;~~\{D,K^\mu \}=K^\mu ~;~~\{K^\mu ,K^\nu \}=0~,
 \label{0s}
\end{equation}
where $J_{\mu\nu}~,T_\mu~,~D$ and $K_\mu $ stand for generators of
Lorentz rotation, translation, dilation and special conformal
transformation respectively. Their structures are given by,
$$
J_{\mu\nu}=X_\mu P_\nu -X_\nu P_\mu~;~~T_\mu = P_\mu ~;~~D=(XP)~;
$$
\begin{equation}
K_\mu =2(XP)X_\mu -X^2P_\mu~.
 \label{1s}
\end{equation}
We exploit the map $(X_\mu ,P_\mu )\rightarrow (x_\mu ,p_\mu )$
given in (\ref{c1}) to rewrite the generators in the $\kappa$-NC
spacetime:
$$
j_{\mu\nu}=x_\mu p_\nu-x_\nu p_\mu ~;~~ t_\mu
=\frac{p_\mu}{1-(\eta p)/\kappa }~;~~~;~~d=(xp);$$
\begin{equation}
k_\mu =(1-(\eta p)/\kappa)[2(xp)x_\mu -x^2p_\mu ]~. \label{s2}
\end{equation}
{\it{By construction, the generators in (\ref{s2}) will satisfy
the same algebra (\ref{0s}) provided one uses the $\kappa$-NC
algebra (\ref{03})}}. These are the deformed generators. The
infinitesimal transformation operators are,
$$
j=\frac{1}{2}a^{\mu\nu}j_{\mu\nu}=\frac{1}{2}a^{\mu\nu}(x_\mu
p_\nu-x_\nu p_\mu )~;~~ t=a^\mu t_\mu =\frac{(ap)}{1-(\eta
p)/\kappa }~;~~d=a(xp)~;$$
\begin{equation}
k=a^\mu k_\mu =(1-(\eta p)/\kappa)[2(xp)(ax) -x^2(ap) ],
\label{s3}
\end{equation}
where generically $a$ denotes the infinitesimal parameter. Using
the definition of small change in $A$ due to transformation
$\delta _b$ as,
\begin{equation}
\delta _b A=\{\delta _b,A\},
 \label{12s}
\end{equation}
 Next we want to ascertain that the $\kappa$-NC
algebra (\ref{03}) is stable under the above symmetry operations.
This is done by checking the validity of the identity,
\begin{equation}
\{A,B\}=C ~\Rightarrow \delta_b \{A,B\}=\delta _bC ,\label{s8}
\end{equation}
or more explicitly,
\begin{equation}
\{\delta_bA,B\}+\{A,\delta_bB\}=\delta _bC .\label{s9}
\end{equation}
In the above we refer to (\ref{03}) for $\{A,B\}=C$ and (\ref{s2})
for $\delta_{b}$. A straightforward but tedious calculation shows
that the above identity is, indeed, valid. This assures us about
the consistency of the whole procedure.
\subsection{Lagrangian for $\kappa$-Particle }
 Lastly we will construct a Lagrangian for the
$\kappa$-particle. This has been a topic of recent interest and
several authors \cite{others}  have proposed models for particles
with NC phase space of different structures. However, the model we
propose here for $\kappa$-NC phase space is quite elegant and can
be expressed in a closed form.

 Again the canonical variable approach becomes
convenient since we are sure that the relativistic free particle
action in terms of canonical $(X_\mu ,P_\mu )$ degrees of freedom
will be,
\begin{equation}
L=(P^\mu\dot X_\mu)-\lambda (P^2-m^2). \label{a1}
\end{equation}
We now convert this $L$ to a function depending on physical
$\kappa $-NC phase space coordinates:
$$
L=(\frac{p^\mu}{1-\frac{(\eta p)}{\kappa }})(x_\mu (1-\frac{(\eta
p)}{\kappa}))^.-\frac{\lambda}{2}(\frac{p^2}{(1-\frac{(\eta
p)}{\kappa})^2}-m^2)$$
\begin{equation}
=(p\dot x)-\frac{(px)(\eta \dot p)}{\kappa (1-\frac{(\eta
p)}{\kappa})}-\frac{\lambda}{2}(p^2-m^2(1-\frac{(\eta
p)}{\kappa})^2), \label{a20}
\end{equation}
where we have redefined the arbitrary multiplier $\lambda $. By
the Hamiltonian constraint analysis, as formulated by Dirac
\cite{dirac} it is straightforward to check  that the symplectic
structure in (\ref{a20}) will induce the $\kappa$-NC phase space
algebra and the $\lambda $-term will obviously impose the MS mass
shell condition.

 Finally, after eliminating the auxiliary variables  we obtain
the cherished Nambu-Goto Lagrangian for the $\kappa $-particle
\cite{gho2},
\begin{equation}
L=\frac{m\sqrt{\dot x^2}}{(1+\frac{m(\eta \dot x)}{\kappa
\sqrt{\dot x^2}})}(1+\frac{m}{\kappa}(\eta \dot x)(\frac{(x\dot
x)}{\sqrt{\dot x^2}})^.). \label{a15}
\end{equation}
Notice that (\ref{a15}) is a higher derivative Lagrangian.
Hamiltonian analysis of it will yield the $\kappa $-NC phase space
algebra.

\section{Deformed and Twisted Gauge Symmetry in Noncommutative Field Theory:}
In this section we consider only canonical noncommutative space -- a space where the noncommutative parameter
$\theta^{\rho\sigma}$  is a real constant antisymmetric
matrix. The functions defined on such a space satisfy the Moyal product (\ref{8}). Replacement of functions by the spacetime coordinates in (\ref{8}), gives
\begin{equation}
[x^{\rho},x^{\sigma}]_{*}=i\theta^{\rho\sigma}. \label{nc}
\end{equation}
This helps one to elevate a commutative field theory to a more
general NC field theory by simply replacing the usual product by
the * product (\ref{8}). Not surprisingly, the NC models defined
on (\ref{nc}) violate the Lorentz invariance just like the algebra
itself. However it has been shown that they are invariant under
the twisted Poincare algebra\cite{cht1}, deformed with the Abelian
twist element
\begin{eqnarray}
\mathcal
F={\textrm{exp}}\left(\frac{i}{2}\theta^{\mu\nu}P_{\mu}P_{\nu}\right)
\label{twist}
\end{eqnarray}
where $P_{\mu}(=\partial_{\mu})$ are the translation generators.
Due to this twist, one can define the following multiplication map for the noncommutative (hatted) variables
\begin{eqnarray}
\mu\circ(\hat \psi\otimes\hat \varphi)=\hat \psi\hat
\varphi\rightarrow\mu_*\circ(\hat \psi\otimes\hat
\varphi)=\mu\circ\mathcal F^{-1}(\hat \psi\otimes\hat
\varphi)\equiv \hat \psi*\hat \varphi
\end{eqnarray}
which is precisely the * product (\ref{8}).

Gauge symmetries can be introduced in these NC models by the usual
gauge invariance requirement. By construction these theories have
the twisted Poincare symmetry and they are invariant under *-gauge
transformations.

Recently an interesting study has been done by twisting not only
the Poincare algebra but also the gauge algebra\cite{vass,Asch}. An
astonishing result was obtained --the NC theory turned out to be
invariant under the commutative space gauge transformation
provided the coproduct of the gauge generators are redefined by
the same twist (\ref{twist}).

Thus, contrary to the commutative theory gauge symmetry in a
noncommutative theory can be interpreted in two different ways. In
one approach star deformed gauge transformations are taken,
keeping the comultiplication (Leibniz) rule unchanged and in the
other approach gauge transformations are taken as in the
commutative case at the expense of a modified Leibniz rule. This
rule is obtained from the usual Leibniz rule by the same twist
operator used in defining the twisted Poincare generators. This
shows a close correspondence between twisted Poincare symmetry and
twisted gauge symmetry. In this review article we analyse both approaches within a
common framework which is a generalization of the treatment of
gauge symmetry in commutative
space\cite{gitman,shirzad,rothe1,rothe2,Hennaux}. We follow both
the Lagrangian and Hamiltonian formulations which are
complementary to each other.

An important point worth mentioning here is that while both types of gauge transformations keep the action invariant some controversies have been raised in the literature. In the case of star gauge transformation, gauge symmetries act only on the fields which is quite analogous to the commutative space theories. Whereas if ordinary gauge transformations with a twisted Leibniz rule is taken, then the transformations do not act only on the fields. Consequently, it is not clear whether the later type of gauge transformation can be considered as a physical symmetry or not. Discussions regarding this issue may be found in \cite{ag,cht}.  
\subsection{Lagrangian Analysis:}
 The model we consider is the NC non-Abelian gauge field coupled with Dirac field{\footnote{Throughout this section we take $\theta^{0i}=0$ to avoid higher order time derivatives.}},
\begin{eqnarray}
S=\int \textrm{d}^4x \ [-\frac{1}{2}\textrm{Tr}({\hat
F}_{\mu\nu}(x)*{\hat F}^{\mu\nu}(x))+{\hat
{\bar{\psi}}}(x)*(i\gamma^{\mu}D_{\mu}*-m){\hat \psi}(x)].
\label{lag}
\end{eqnarray}
where
\begin{eqnarray}
D_{\mu}*\hat{\psi}(x)&\equiv&\partial_{\mu}+ig{\hat A}_{\mu}*\hat{\psi}(x)\\
{\hat F}_{\mu\nu}(x)&\equiv&\partial_{\mu}{\hat
A}_{\nu}(x)-\partial_{\nu}{\hat A}_{\mu}(x)+ig[{\hat
A}_{\mu}(x),{\hat A}_{\nu}(x)]_*. \label{f}
\end{eqnarray}
The action (\ref{lag}) is invariant under both {\it{deformed gauge
transformations}} \cite{sw},
\begin{eqnarray}
&&\delta {\hat A}_{\mu}=\mathcal{D}_{\mu}*\hat\alpha=\partial_{\mu}\hat\alpha+ig({\hat A}_{\mu}*\hat\alpha-\hat\alpha*{\hat A}_{\mu}),\label{Amu}\\
&&\delta {\hat F}_{\mu\nu}=ig[{\hat F}_{\mu\nu},\hat\alpha]_*=ig({\hat F}_{\mu\nu}*\hat\alpha-\hat\alpha*{\hat F}_{\mu\nu})\label{Fmunu}\\
&&\delta {\hat \psi}=-ig\hat\alpha*{\hat \psi}\label{si}\\
&&\delta
{\hat{\bar{\psi}}}=ig{\hat{\bar{\psi}}}*\hat\alpha\label{sibar}
\end{eqnarray}
with the {\it{usual Leibniz Rule}} (LR),
\begin{eqnarray}
\delta (f*g)=(\delta f)*g+f*(\delta g) \label{tX}
\end{eqnarray}
as well as the {\it{undeformed gauge transformations}}
\begin{eqnarray}
&&\delta_{\hat\alpha} {\hat A}_{\mu}=\mathcal{D}_{\mu}\hat\alpha=\partial_{\mu}\hat\alpha+ig({\hat A}_{\mu}\hat\alpha-\hat\alpha {\hat A}_{\mu}),\label{YY1}\\
&&\delta_{\hat\alpha} {\hat F}_{\mu\nu}=ig[{\hat F}_{\mu\nu},\hat\alpha]=ig({\hat F}_{\mu\nu}\hat\alpha-\hat\alpha {\hat F}_{\mu\nu})\\
&&\delta_{\hat\alpha} \hat \psi=-ig\hat\alpha\hat \psi\\
&&\delta_{\hat\alpha}
{\hat{\bar{\psi}}}=ig{\hat{\bar{\psi}}}\hat\alpha
\label{YY}
\end{eqnarray}
with the {\it{twisted Leibniz Rule}} (TLR)\cite{vass,Asch,jWess},
\begin{eqnarray}
\delta_{\hat\alpha}(f*g)&=&\sum_n(\frac{-i}{2})^n\frac{\theta^{\mu_1\nu_1}\cdot \cdot \cdot\theta^{\mu_n\nu_n}}{n!}\nonumber\\
&&(\delta_{\partial_{\mu_1}\cdot \cdot
\cdot\partial_{\mu_n}\hat\alpha}f*\partial_{\nu_1}\cdot \cdot
\cdot\partial_{\nu_n}g+\partial_{\mu_1}\cdot \cdot
\cdot\partial_{\mu_n}f*\delta_{\partial_{\nu_1}\cdot \cdot
\cdot\partial_{\nu_n}\hat\alpha}g). \label{co}
\end{eqnarray}
We now elaborate on the derivation of the above rule (\ref{co}), following Aschieri et. al.\cite{Asch}. the *-product of two functions is defined in terms of a twist operator $\mathcal{F}_{\theta}(=e^{\frac{i}{2}\theta^{\rho\sigma}\partial_{\rho}\otimes\partial_{\sigma}})$
\begin{eqnarray}
f*g=m_0\{\mathcal{F}_{\theta}f\otimes g\}
\end{eqnarray}
where the map $m_0$ has been defined in (\ref{9}). The twist operator can now be inverted to write the ordinary product as,
\begin{eqnarray}
f\cdot g=\left(\sum_{n=0}^{\infty}(-\frac{i}{2})^n\frac{1}{n!}\theta^{\rho_1\sigma_1}...\theta^{\rho_n\sigma_n}(\partial_{\rho_1}...\partial_{\rho_n}f)*\partial^*_{\sigma_1}...\partial^*_{\sigma_n}f\right)*g\label{qwer}
\end{eqnarray}
where $\partial_{\rho}^* $ is defined in the following way
\begin{eqnarray}
\partial_{\rho}^* : \  \  \  \ && \partial_{\rho}^* f\equiv \partial_{\rho}f\\
&& \partial_{\rho}^*(f*g)= (\partial_{\rho}^* f)*g + f*(\partial_{\rho}^* g)
\end{eqnarray}
The equation (\ref{qwer}) shows that $f\cdot g$ can be viewed as the *-action of a differential operator $X_f^*$ on $g$ 
\begin{eqnarray}
f\cdot g= X_f^**g
\end{eqnarray}
where
\begin{eqnarray}
X_f^*=\sum_{n=0}^{\infty}(-\frac{i}{2})^n\frac{1}{n!}\theta^{\rho_1\sigma_1}...\theta^{\rho_n\sigma_n}(\partial_{\rho_1}...\partial_{\rho_n}f)*\partial^*_{\sigma_1}...\partial^*_{\sigma_n}f
\end{eqnarray}
Since the *-product is associative one can show that
\begin{eqnarray}
X_f^** X_g^*= X_{f\cdot g}^*
\end{eqnarray}
Now a gauge transformation
\begin{eqnarray}
\delta_{\alpha}\psi(x)=i\alpha\cdot\psi=i\alpha^a(x)T^a\psi(x)
\end{eqnarray}
can be interpreted as a *-action
\begin{eqnarray}
\delta_{\alpha}\psi=iX_{\alpha^a}^**T^a\psi=iX_{\alpha}^**\psi=i\alpha\cdot\psi
\end{eqnarray}
In a commutative space gauge theory the Hopf algebra
\begin{eqnarray}
\Delta\delta_{\alpha}(\phi\otimes\psi)=(\delta_{\alpha}\phi)\otimes\psi+\phi\otimes(\delta_{\alpha}\psi)
\end{eqnarray}
together with the gauge transformations of the basic fields give the transformation of the product of fields
\begin{eqnarray}
\delta_{\alpha}(\phi\cdot\psi)= \delta_{\alpha}m_0(\phi\otimes\psi)=m_0 \Delta\delta_{\alpha} (\phi\otimes\psi)
\end{eqnarray}
In order to extend a Lie algebra to a Hopf algebra in noncommutative space the operator $ \mathcal{F}_{\theta}^{-1}$ can be used in a convenient manner to get the following coproduct
\begin{eqnarray}
\Delta_{\theta}(\delta_{\hat{\alpha}})(\phi\otimes\psi)=i\mathcal{F}_{\theta}^{-1}(\hat{\alpha}\otimes 1+1\otimes\hat{\alpha})\mathcal{F}_{\theta} (\phi\otimes\psi)
\end{eqnarray}
which when acted by $m_0$ gives the expression (\ref{co}).

The former (LR) formalism is probably more familiar and the latter
(TLR) is a relatively new development \cite{cht,cht1,rb}. In our
analysis we take ${\hat A}_{\mu}$ and ${\hat F}_{\mu\nu}$ to be
enveloping algebra valued, {\it{i.e.}} they are expanded over the
basis $T^a$ which satisfy the Lie algebraic relation
\begin{eqnarray}
[T^a,T^b]=if^{abc}T^c ~~;~~ \{T^a,T^b\}=d^{abc}T^c. \label{dabc}
\end{eqnarray}
We also impose the trace condition
\begin{eqnarray}
{\textrm{Tr}}(T^aT^b)=\frac{1}{2}\delta^{ab} \label{trace}
\end{eqnarray}
so that $f^{abc}$ is completely antisymmetric and $d^{abc}$ is completely symmetric.\\
{Leibniz Rule \it{(LR) formalism}}:  We start by discussing the LR
formalism. Let us recall that if $L_{a}$ denotes the Euler
derivatives for a general Lagrangian density
$\mathcal{L}\left(q_{a}, \
\partial_{\mu}q_{a},\right)$, corresponding to each gauge symmetry
of the Lagrangian  an identity can be written in terms of $L_{a}$
as\cite{gitman,shirzad,samanta}
\begin{eqnarray}
\Lambda^b({\bf{z}},t)=\left[\sum_{s=0}^n\int \textrm{d}^3{\bf{x}}
\ \frac{\partial^s}{\partial t^s}\left(\rho^{a
b}_{(s)}(x,z)L_{a}({\bf{x}},t)\right)\right]=0. \label{lam}
\end{eqnarray}
where gauge transformation of the fields are of the form
\begin{eqnarray}
\delta
q^{a}({\bf{x}},t)=\sum_{s=0}^n(-1)^s\int\textrm{d}^3{\bf{z}} \
\frac{\partial^s\alpha^b({\bf{z}},t)}{\partial t^s}\rho^{a
b}_{(s)}(x,z) \label{a}
\end{eqnarray}
with $\alpha$ and $\rho$ being the parameter and generator
respectively, of the transformation.

We show explicitly how to derive the deformed gauge transformation
rules (\ref{Amu}-\ref{sibar}). The equations of motion for the
action (\ref{lag}) can be obtained by setting Euler derivatives to
zero. These are given by
\begin{eqnarray}
&&L^{\mu a}=-\left(\mathcal{D}_{\sigma}*{\hat
F}^{\sigma\mu}\right)^a-g\hat\psi_j(\gamma^{\mu}T^a)_{ij}*{\hat{\bar{\psi}}}_i
\label{eu1}\\
&&L_i=-i\partial_{\mu}{\hat{\bar{\psi}}}_j(\gamma^{\mu})_{ji}-g{\hat{\bar{\psi}}}_j*(\gamma^{\mu}{\hat
A}_{\mu}^aT^a)_{ji}-m{\hat{\bar{\psi}}}_i
\label{eu2}\\
&&L_i'=-i(\gamma^{\mu})_{ij}\partial_{\mu}{\hat\psi}_j+g(\gamma^{\mu}{\hat
A}_{\mu}^aT^a)_{ij}*{\hat\psi}_j+m\hat\psi_i. \label{eu3}
\end{eqnarray}
Here the noncommutative covariant derivative $\mathcal{D}*$ is
defined in the adjoint representation (\ref{Amu}). The gauge
identity for this system is \cite{rb,samanta1}
\begin{eqnarray}
\Lambda^a\equiv-\left(\mathcal{D}^{\mu}*L_{\mu}\right)^a-igT^a_{ij}{\hat\psi}_j*L_i-igT^a_{ji}L_i'*{\hat{\bar{\psi}}}_j=0.
\label{lambda}
\end{eqnarray}
 Comparing (\ref{lambda}) and (\ref{lam}) different gauge generators can be obtained. As an example,
\begin{eqnarray}
\Lambda^a|_{L_i}=\frac{g}{2}f^{abc}\{{\hat
A}^{ib},L_i^c\}_*-i\frac{g}{2}d^{abc}[{\hat
A}^{ib},L_i^c]_*-\partial^{iz}L_{i}^a. \label{ku}
\end{eqnarray}

Using the properties
\begin{eqnarray}
\int \textrm{d}^4 x \ A(x)*B(x)=\int \textrm{d}^4 x \
A(x)B(x)=\int \textrm{d}^4x \ B(x)*A(x) \label{b1}
\end{eqnarray}
and
\begin{eqnarray}
\int \textrm{d}^4 x \ (A*B*C)=\int \textrm{d}^4 x \
(B*C*A)=\int\textrm{d}^4 x \ (C*A*B) \label{b2}
\end{eqnarray}
(\ref{ku}) is written in the following way
\begin{eqnarray}
&&\Lambda^a|_{L_i}({\bf{z}},t)\nonumber\\
&=&-\int \textrm{d}^3{\bf{x}} \ \frac{g}{2}\left(f^{abc}\{\delta^3({\bf{x}}-{\bf{z}}),{\hat A}^{ic}(x)\}_*+id^{abc}[\delta^3({\bf{x}}-{\bf{z}}),{\hat A}^{ic}(x)]_*\right)*L_i^b(x)\nonumber\\
&&-\int \textrm{d}^3{\bf{x}} \
\delta^{ab}\partial^{i{\bf{z}}}\delta^3({\bf{x}}-{\bf{z}})L_{i}^b(x).
\label{bdr}
\end{eqnarray}
Comparing with (\ref{lam}) we obtain,
\begin{eqnarray}
\rho^{bia}_{(0)}(x,z)&=&-\delta^{ab}\partial^{i{\bf{z}}}\delta^3({\bf{x}}-{\bf{z}})-\nonumber\\
 &&\frac{g}{2}f^{abc}\{\delta^3({\bf{x}}-{\bf{z}}),{\hat A}^{ic}(x)\}_*-i\frac{g}{2}d^{abc}[\delta^3({\bf{x}}-{\bf{z}}),{\hat A}^{ic}(x)]_*.
\label{37}
\end{eqnarray}
Other components of the gauge generator can be obtained in a
similar way. These are
\begin{eqnarray}
\rho^{b0a}_{(0)}(x,z)&=&-\frac{g}{2}f^{abc}\{\delta^3({\bf{x}}-{\bf{z}}),{\hat
A}_0^c(x)\}_*-i\frac{g}{2}d^{abc}[\delta^3({\bf{x}}-{\bf{z}}),{\hat
A}_0^c(x)]_*
\label{a10}\\
\rho^{b0a}_{(1)}(x,z)&=&-\delta^{ab}\delta^3({\bf{x}}-{\bf{z}})
\label{a2}\\
\phi^a_{i(0)}(x,z)&=&-igT^a_{ij}\delta^3({\bf{x}}-{\bf{z}})*\hat\psi_j(x)\\
\phi'^a_{i(0)}(x,z)&=&-igT^a_{ji}{\hat{\bar{\psi}}}_j(x)*\delta^3({\bf{x}}-{\bf{z}})
\end{eqnarray}

Let us next consider the gauge transformations. From (\ref{a}) we
write the gauge transformation equation for the space component of
the gauge field
\begin{eqnarray}
\delta {\hat A}^{i a}({\bf{x}},t)&=&\sum_s(-1)^s\int\textrm{d}^3{\bf{z}} \ \frac{\partial^s\hat\alpha^b({\bf{z}},t)}{\partial t^s}*\rho^{aib}_{(s)}(x,z)\nonumber\\
&=&\int\textrm{d}^3{\bf{z}} \
\left(\hat\alpha^b({\bf{z}},t)*\rho^{aib}_{(0)}(x,z)\right)
\label{mono}
\end{eqnarray}
Exploiting the identity \cite{amorim,rb,samanta1}
\begin{eqnarray}
A(x)*\delta(x-z)=\delta(x-z)*A(z) \label{delta*}
\end{eqnarray}
 and interchanging $a, \ b$, the generator (\ref{37}) is recast as,
\begin{eqnarray}
\rho^{aib}_{(0)}(x,z)&=&-\delta^{ab}\partial^{i{\bf{z}}}\delta^3({\bf{x}}-{\bf{z}})+\nonumber\\
 &&\frac{g}{2}f^{abc}\{\delta^3({\bf{x}}-{\bf{z}}),{\hat A}^{ic}(z)\}_*+i\frac{g}{2}d^{abc}[\delta^3({\bf{x}}-{\bf{z}}),{\hat A}^{ic}(z)]_*
\label{a3}
\end{eqnarray}

Use of (\ref{a3}) along with the identities (\ref{b1}) and
(\ref{b2}) in (\ref{mono}) implies that
\begin{eqnarray}
\delta {\hat
A}^{ia}=\partial^{i}\hat\alpha^{a}-\frac{g}{2}f^{abc}\{{\hat
A}^{ib},\hat\alpha^{c}\}_*+i\frac{g}{2}d^{abc}[{\hat
A}^{ib},\hat\alpha^{c}]_*=(\mathcal{D}^{i}*\hat\alpha)^a
\label{r1}
\end{eqnarray}
This is the space component of (\ref{Amu}). The time component can also be obtained in a similar manner\cite{samanta1}.\\
{Twisted Leibniz Rule \it{TLR formalism}}: Now we show how the TRL
appears. For simplicity we take the pure gauge theory
\begin{eqnarray}
S=-\frac{1}{2}\int \textrm{d}^4x \ {\textrm {Tr}}({\hat
F}_{\mu\nu}(x)*{\hat F}^{\mu\nu}(x)) \label{s}
\end{eqnarray}
Using the twisted gauge transformation (\ref{co})
\begin{eqnarray}
\delta_{\hat\alpha}({\hat A}_{\mu}*{\hat
A}_{\nu})=\partial_{\mu}\hat\alpha {\hat A}_{\nu}+{\hat
A}_{\mu}\partial_{\nu}\hat\alpha-ig[\hat\alpha,({\hat
A}_{\mu}*{\hat A}_{\nu})] \label{gag}
\end{eqnarray}
the gauge transformation of the field strength tensor
\begin{eqnarray}
\delta_{\hat\alpha}{\hat F}_{\mu\nu}&=&\partial_{\mu}\delta_{\hat\alpha}{\hat A}_{\nu}-\partial_{\nu}\delta_{\hat\alpha}{\hat A}_{\nu}+ig\delta_{\hat\alpha}[{\hat A}_{\mu},{\hat A}_{\nu}]_*\\
&=&\partial_{\mu}(\partial_{\nu}\hat\alpha+ig[{\hat
A}_{\nu},\hat\alpha])-\partial_{\nu}(\partial_{\mu}\hat\alpha+ig[{\hat
A}_{\mu},\hat\alpha])\nonumber\\&&+ig\left([\partial_{\mu}\hat\alpha,{\hat
A}_{\nu}]+[{\hat
A}_{\mu},\partial_{\nu}\hat\alpha]-ig[\hat\alpha,[{\hat
A}_{\mu},{\hat A}_{\nu}]_*]\right)
\label{GF}\\
&=&-ig[\hat\alpha,{\hat F}_{\mu\nu}].
\end{eqnarray}
Likewise one finds,
\begin{eqnarray}
\delta_{\hat\alpha}({\hat F}^{\mu\nu}*{\hat
F}_{\mu\nu})=-ig[\hat\alpha,{\hat F}^{\mu\nu}*{\hat F}_{\mu\nu}]
\end{eqnarray}
Both ${\hat F}_{\mu\nu}$ and ${\hat F}_{\mu\nu}*{\hat F}^{\mu\nu}$ have the usual
(undeformed) transformation properties. Thus the action (\ref{s}) is invariant under the
gauge transformation (\ref{gag}) and the deformed coproduct rule (\ref{co}).\\
 There is another way of interpreting the gauge invariance which makes contact with the
gauge identity.

Making a gauge variation of the action (\ref{s})
and taking into account the twisted coproduct rule (\ref{co}), we
get
\begin{eqnarray}
\delta_{\hat\alpha}S&=&-\frac{1}{2}\int \textrm{d}^4x \ {\textrm {Tr}}\delta_{\hat\alpha}({\hat F}_{\mu\nu}*{\hat F}^{\mu\nu})\\
&=&-\frac{1}{2}\int \textrm{d}^4x \ [{\textrm
{Tr}}(\delta_{\hat\alpha}{\hat F}_{\mu\nu}*{\hat F}^{\mu\nu}+{\hat
F}_{\mu\nu}*\delta_{\hat\alpha}{\hat
F}^{\mu\nu}\nonumber\\&&-\frac{i}{2}\theta^{\mu_1\nu_1}(\delta_{\partial_{\mu_1}\hat\alpha}{\hat
F}_{\mu\nu}*\partial_{\nu_1}{\hat
F}^{\mu\nu}+\partial_{\mu_1}{\hat
F}_{\mu\nu}*\delta_{\partial_{\nu_1}\hat\alpha}{\hat
F}^{\mu\nu})\nonumber\\&&+\cdot\cdot\cdot)]. \label{delS}
\end{eqnarray}
Now using the result (\ref{GF}) each term of (\ref{delS}) can be
computed separately. For example we concentrate on the first term.
Using the identity (\ref{b1}) and the trace condition
(\ref{trace}) we write the first term as
\begin{eqnarray}
\delta_{\hat\alpha}S|_{{\textrm{1st term}}}&=&-\frac{1}{4}\int \textrm{d}^4x \ (\delta_{\hat\alpha}{\hat F}^{\mu\nu a}*{\hat F}_{\mu\nu}^a+{\hat F}^{\mu\nu a}*\delta_{\hat\alpha}{\hat F}_{\mu\nu}^a)\\
&=&-\frac{1}{2}\int \textrm{d}^4x \ \delta_{\hat\alpha}{\hat
F}^{\mu\nu a}{\hat F}_{\mu\nu}^a.
\end{eqnarray}
Making use of (\ref{GF}) and dropping the surface terms the above
expression is found out to be,
\begin{eqnarray}
\delta_{\hat\alpha}S|_{{\textrm{1st term}}}=-\int
\textrm{d}^4x&\hat\alpha^a&(-\partial^{\mu}\partial^{\nu}{\hat
F}_{\mu\nu}-ig\partial^{\mu}[{\hat A}^{\nu},{\hat
F}_{\mu\nu}]-ig[{\hat A}^{\mu},\partial^{\nu}{\hat
F}_{\mu\nu}]\nonumber\\&&+g^2[{\hat A}^{\mu}*{\hat A}^{\nu},{\hat
F}_{\mu\nu}])^a.
\end{eqnarray}
The second term of (\ref{delS}) is identically zero due to the
antisymmetric nature of $\theta^{\mu\nu}$. We write that as,
\begin{eqnarray}
\delta_{\hat\alpha}S|_{{\textrm{2nd term}}}&=&-\frac{1}{2}\int \textrm{d}^4x \ \hat\alpha^a\frac{i}{2}\theta^{\mu_1\nu_1}(-ig\{\partial_{\mu_1}{\hat F}^{\mu\nu},\partial_{\nu_1}{\hat F}_{\mu\nu}\})^a\\
&=&-\int \textrm{d}^4x \ \hat\alpha^a\frac{i}{2}\theta^{\mu_1\nu_1}(-ig\{\partial_{\mu_1}\partial^{\mu}{\hat A}^{\nu},\partial_{\nu_1}{\hat F}_{\mu\nu}\}\nonumber\\&&+g^2\{\partial_{\mu_1}({\hat A}^{\mu}*{\hat A}^{\nu}),\partial_{\nu_1}{\hat F}_{\mu\nu}\})^a\\
&=&-\int \textrm{d}^4x \
\hat\alpha^a\frac{i}{2}\theta^{\mu_1\nu_1}(-ig\partial^{\mu}\{\partial_{\mu_1}{\hat
A}^{\nu},\partial_{\nu_1}{\hat
F}_{\mu\nu}\}-\nonumber\\&&ig\{\partial_{\mu_1}{\hat
A}^{\mu},\partial_{\nu_1}\partial^{\nu}{\hat
F}_{\mu\nu}\}+g^2\{\partial_{\mu_1}({\hat A}^{\mu}*{\hat
A}^{\nu}),\partial_{\nu_1}{\hat F}_{\mu\nu}\})^a.
\end{eqnarray}
Other terms can be obtained in a similar manner. Combining all
these terms we finally get,
\begin{eqnarray}
\delta_{\hat\alpha}S&=&-\int \textrm{d}^4x \ \hat\alpha^a(-\partial^{\mu}\partial^{\nu}{\hat F}_{\mu\nu}-ig\partial^{\mu}[{\hat A}^{\nu},{\hat F}_{\mu\nu}]_*-ig[{\hat A}^{\mu},\partial^{\nu}{\hat F}_{\mu\nu}]_*\nonumber\\&&+g^2[{\hat A}^{\mu}*{\hat A}^{\nu},{\hat F}_{\mu\nu}]_*)^a\\
&=&-\int \textrm{d}^4x \ \hat\alpha^a\Lambda^a
\end{eqnarray}
where,
\begin{eqnarray}
\Lambda^a=-(\mathcal{D}^{\mu}*L_{\mu})^a=-(\mathcal{D}^{\mu}*\mathcal{D}^{\sigma}*{\hat
F}_{\sigma \mu})^a \label{70}
\end{eqnarray}
that vanishes identically. Note that this is exactly the same as
the expression in the gauge identity (\ref{lambda}) without the
fermionic fields. This proves the invariance of the action.

Let us now repeat the analysis of the previous section for TLR.
Since the gauge transformations are undeformed, the gauge
generators are expected to have the same form as in the
commutative space. To see this note that the gauge variation of
the zeroth component of the ${\hat A}_{\mu}$ field, following from
(\ref{YY1}), can be written as,
\begin{eqnarray}
\delta_{\hat\alpha}{\hat
A}_0^a(z)&=&\partial_0\hat\alpha^a(z)-gf^{abc}{\hat
A}_0^b(z)\hat\alpha^c(z)\cr &=&g\int \textrm{d}^3{\bf{z}} \
f^{abc}{\hat
A}_0^c\hat\alpha^b\delta^3({\bf{x}}-{\bf{z}})+\int\textrm{d}^3{\bf{z}}
\ \delta^{ab}\delta^3({\bf{x}}-{\bf{z}})\frac{\partial}{\partial
t}\hat\alpha^b.
\end{eqnarray}
Clearly the above result can be expressed in our standard  form
(\ref{a}),
\begin{eqnarray}
\delta_{\hat\alpha} {\hat
A}_0^a(z)&=&\sum_s(-1)^s\int\textrm{d}^3{\bf{z}}\frac{\partial^s\hat\alpha^b({\bf{z}},t)}
{\partial t^s}\rho^{a0b}_{(s)}(x,z)\cr &=&\int\textrm{d}^3{\bf{z}}
\
\hat\alpha^b({\bf{z}},t)\rho^{a0b}_{(0)}(x,z)-\int\textrm{d}^3{\bf{z}}
\ \frac{\partial \hat\alpha^b({\bf{z}},t)}{\partial
t}\rho^{a0b}_{(1)}(x,z)
\end{eqnarray}
where
\begin{eqnarray}
&&\rho^{a0b}_{(0)}(x,z)=gf^{abc}{\hat
A}_0^c\delta^3({\bf{x}}-{\bf{z}})
\label{r11}\\
&&\rho^{a0b}_{(1)}(x,z)=-\delta^{ab}\delta^3({\bf{x}}-{\bf{z}})
\label{r22}
\end{eqnarray}
is the gauge generator. Similarly from the space part of
(\ref{YY1}) we find
\begin{eqnarray}
\rho^{aib}_{(0)}(x,z)=-\delta^{ab}\partial^{i{\bf{z}}}\delta^3({\bf{x}}-{\bf{z}})+gf^{abc}{\hat
A}_i^c\delta^3({\bf{x}}-{\bf{z}}). \label{r33}
\end{eqnarray}
Now as already implied in (\ref{70}), there is a gauge identity
for this system,\begin{eqnarray}
\Lambda^a=-\left(\mathcal{D}^{\mu}*L_{\mu}\right)^a=0 \label{su}
\end{eqnarray}
where $L_{\mu}$ is the Euler derivative defined in (\ref{70}). The
gauge identity and the Euler derivatives are mapped by the
relation \cite{rb,samanta1},
\begin{eqnarray}
\Lambda^a({\bf{z}},t)=\sum_{s=0}^n\int \textrm{d}^3{\bf{x}} \
\frac{\partial^s}{\partial t^s}\left(\rho'^{b\mu
a}_{(s)}(x,z)L^b_{\mu}({\bf{x}},t)\right)
\end{eqnarray}
where the values of $\rho'^{b\mu a}_{(0)}(x,z)$ and $\rho'^{b\mu
a}_{(1)}(x,z)$ are equal to those of $\rho^{b\mu a}_{(0)}$ and
$\rho^{b\mu a}_{(1)}$ of the previous example, given in
(\ref{a10}), (\ref{a2}) and (\ref{37}). This happens since the
Euler derivatives and the gauge identity are identical to those
discussed above. Now we express $\rho'$ in terms
of $\rho$. To do this, (\ref{a1}) is rewritten under the
identification $\rho=\rho'$ as,
\begin{eqnarray}
\rho'^{b0a}_{(0)}(x,z)=-\frac{g}{2}f^{abc}\{\delta^3({\bf{x}}-{\bf{z}}),{\hat
A}_0^c(x)\}_*-i\frac{g}{2}d^{abc}[\delta^3({\bf{x}}-{\bf{z}}),{\hat
A}_0^c(x)]_*.
\end{eqnarray}
Now making use of the definition of star product, the above
expression is written in the following way
\begin{eqnarray}
\rho'^{b0a}_{(0)}(x,z)&=&-gf^{abc}{\hat A}^c_0\delta^3({\bf{x}}-{\bf{z}})-g\sum_{n=1}^{\infty}(\frac{i}{2})^n\frac{\theta^{\mu_1\nu_1}\cdot\cdot\cdot\theta^{\mu_n\nu_n}}{n!}\nonumber\\
&&[(\frac{f^{abc}}{2}+i\frac{d^{abc}}{2})\partial_{\mu_1}\cdot\cdot\cdot\partial_{\mu_n}\delta^3({\bf{x}}-{\bf{z}})\partial_{\nu_1}\cdot\cdot\cdot\partial_{\nu_n}{\hat A}^{0c}(x)\\
&&(+\frac{f^{abc}}{2}-i\frac{d^{abc}}{2})
\partial_{\mu_1}\cdot\cdot\cdot\partial_{\mu_n}{\hat A}^{0c}(x)\partial_{\nu_1}\cdot\cdot\cdot\partial_{\nu_n}\delta^3({\bf{x}}-{\bf{z}})].\nonumber
\end{eqnarray}
Note that the $\theta$ independent term is nothing but the gauge
generator $\rho^{b0a}_{(0)}$ (or -$\rho^{a0b}_{(0)}$) given in (\ref{r11}). Similarly
calculating the other components $\rho'^{bia}_{(0)}$ and
$\rho'^{b0a}_{(1)}$ from (\ref{a2}) and (\ref{37}) we obtain,
\begin{eqnarray}
\rho'^{b\mu a}_{(0)}(x,z)&=&\rho^{b\mu a}_{(0)}(x,z)-g\sum_{n=1}^{\infty}(\frac{i}{2})^n\frac{\theta^{\mu_1\nu_1}\cdot\cdot\cdot\theta^{\mu_n\nu_n}}{n!}\nonumber\\
&&[(\frac{f^{abc}}{2}+i\frac{d^{abc}}{2})\partial_{\mu_1}\cdot\cdot\cdot\partial_{\mu_n}\delta^3({\bf{x}}-{\bf{z}})\partial_{\nu_1}\cdot\cdot\cdot\partial_{\nu_n}{\hat
A}^{\mu c}(x)
\label{tui}\\
&&(+\frac{f^{abc}}{2}-i\frac{d^{abc}}{2})
\partial_{\mu_1}\cdot\cdot\cdot\partial_{\mu_n}{\hat A}^{\mu c}(x)\partial_{\nu_1}\cdot\cdot\cdot\partial_{\nu_n}\delta^3({\bf{x}}-{\bf{z}})]\nonumber\\
\rho'^{b0a}_{(1)}(x,z)&=&\rho^{b0a}_{(1)}(x,z).
\end{eqnarray}

Here the generator of the system is $\rho$
(\ref{r11},\ref{r22},\ref{r33}). Although it remains undeformed,
the relation mapping the gauge identity with the generator is
twisted. The additional twisted terms are explicitly given in
(\ref{tui}). In the commutative space limit the twisted terms
naturally vanish.

\subsection{Hamiltonian Analysis}
We consider a system with a canonical Hamiltonian $H_c$ and a set
of first class constraints $\Phi_{a}\approx0$. In general
$\Phi_{a}$ includes both the primary ($\Phi_{a_1}$) and secondary
constraints ($\Phi_{a_2}$) and satisfy the following involutive
Poisson algebra \cite{gitman,rothe1,rothe2,Hennaux,banerjee22,dirac}

\begin{eqnarray}
&&\{H_c,\Phi_a(x)\}=\int \textrm{d}y \ V^b_a(x,y)\Phi_b(y),
\label{ve}\\
&&\{\Phi_a(x),\Phi_b(y)\}=\int \textrm{d}z \
C^c_{ab}(x,y,z)\Phi_c(z) \label{cc}
\end{eqnarray}
where $V$ and $C$ are structure functions. The gauge
transformation of a variable $F$ is obtained from the Poisson
bracket
\begin{eqnarray}
\delta F(x) =\int \textrm{d}y \  \{F(x),G(y)\}. \label{gt}
\end{eqnarray}
where $G$ is the generator of the system. According to Dirac's
algorithm it is a linear combination of all the first class
constraints,
\begin{eqnarray}
&&G=\int \textrm{d}x \ \epsilon^a(x)*\Phi_a(x). \label{G6}
\end{eqnarray}
Here the number of independent gauge parameter is $a_1$. Other
parameters are fixed by the relation\cite{rothe1,rothe2}
\begin{eqnarray}
\frac{\textrm{d}\epsilon^{b_2}(x)}{\textrm{d}t}&=&\int \textrm{d}y \ \epsilon^a(y)V^{b_2}_a(y,x)\nonumber\\
&&+\int \textrm{d}y \ \textrm{d}z \
\epsilon^a(y)v^{a_1}(z)C^{b_2}_{a_1a}(z,y,x) \label{sl}
\end{eqnarray}
{\it{LR formalism}}: The general analysis discussed above is now
used here for the model (\ref{lag}) to study its Hamiltonian
description. Due to the presence of grassmanian variables in our
model (\ref{lag}), the Poisson brackets should be replaced by the
graded brackets\footnote{For fermions
$\{\hat\psi_{\alpha}(x),{\hat\psi}_{\beta}^{\dagger}(y)\}=-i\delta_{\alpha\beta}\delta(x-y)$.}.
The canonical momenta of the Lagrangian (\ref{lag}),
\begin{eqnarray}
\hat{\pi}^a_{\sigma}=\frac{\partial\mathcal{L}}{\partial
\dot{{\hat A}}^{\sigma a}}=\hat{F}^a_{\sigma 0} \label{mmomenta}
\end{eqnarray}
satisfy the basic Poisson bracket relation
\begin{eqnarray}
\{{\hat A}^{\mu
a}(x),\hat{\pi}^b_{\nu}(y)\}=\delta^{ab}\delta^{\mu}_{\nu}\delta(x-y)
\label{bbracket}
\end{eqnarray}
The zeroth component of the momenta (\ref{mmomenta}) leads to a
primary constraint
\begin{eqnarray}
\Phi_1^a=\hat{\pi}^a_0\approx0. \label{con1001}
\end{eqnarray}
The canonical Hamiltonian of the system is given by,
\begin{eqnarray}
H&=&\int \textrm{d}x \  [\frac{1}{2}\hat{\pi}^{ic}*\hat{\pi}^{ic}+\frac{1}{4}{\hat F}_{ij}^a*{\hat F}^{ija}-(\mathcal{D}_i*\hat{\pi}^i)^a*{\hat A}_0^a\nonumber\\
&&-i{\hat{\bar{\psi}}}*\gamma^i\partial_i\hat\psi+g{\hat{\bar{\psi}}}*\gamma^{\mu}{\hat
A}_{\mu}*\hat\psi+m{\hat{\bar{\psi}}}*\hat\psi] \label{ham001}
\end{eqnarray}
Now  using (\ref{bbracket}), the secondary constraints of the
system are computed
\begin{eqnarray}
&&\Phi_2^a=\{H,\Phi_1^a\}=\{H,\hat{\pi}^a_0\}=(\mathcal{D}_i*\hat{\pi}_i)^a-g\hat\psi_{\lambda}*
(T^a)_{\sigma\lambda}(\hat\psi^{\dagger})_{\sigma}\approx 0.
\label{con2001}
\end{eqnarray}
Note that this constraint is the zeroth component of the equation
of motion of the gauge field expressed in phase space variables.
The constraint algebra found is \cite{amorim}
\begin{eqnarray}
\{\Phi^a_2(x),\Phi^b_2(y)\}=\frac{g}{2}f^{abc}\{\delta(x-y),\Phi^c_2(x)\}_*-i\frac{g}{2}d^{abc}[\delta(x-y),\Phi^c_2(x)]_*.
\end{eqnarray}
All other brackets are zero. The involutive algebra of the
canonical Hamiltonian with the constraints is found to be,
\begin{eqnarray}
&&\{H_c,\Phi_1^a\}=\Phi_2^a
\label{al}\\
&&\{H_c,\Phi_2^a\}=-\frac{g}{2}f^{abc}\{{\hat
A}^{0b},\Phi_2^c\}_*+i\frac{g}{2}d^{abc}[{\hat
A}^{0b},\Phi_2^c]_*. \label{al1}
\end{eqnarray}
The $V$ function defined in (\ref{ve}) is found from the algebra
(\ref{al}) and (\ref{al1})
\begin{eqnarray}
(V^2_1)^{ab}(x,y)&=&\delta^{ab}\delta(x-y),
\label{50}\\
(V^2_2)^{ab}(x,y)&=&\frac{g}{2}f^{abc}\{\delta(x-y),{\hat A}^{0c}(y)\}_*\nonumber\\
&&+i\frac{g}{2}d^{abc}[\delta(x-y),{\hat A}^{0c}(y)]_*.
\label{510}
\end{eqnarray}
Now the term $C^{b_2}_{a_1a}$ of (\ref{cc}) vanishes due to the
constraint algebra. So from (\ref{sl}) we get
$\epsilon^{1}=(\mathcal{D}_0*\epsilon^2)$\cite{rb,banerjee22}.
Thus (\ref{G6}) is expressed in terms of the single parameter
($\epsilon^2$) as,
\begin{eqnarray}
G=\int \textrm{d}x \
(\mathcal{D}_0*\epsilon^{2})^a*\Phi_1^a+\epsilon^{2a}*\Phi_2^a
\label{generator}
\end{eqnarray}
where the constraints $\Phi_1$ and $\Phi_2$ were defined in
(\ref{con1001}) and (\ref{con2001}). After obtaining the complete
form of the generator, we can now calculate the variation of the
different fields from (\ref{gt}),

Let us first study the gauge transformation of the field ${\hat
A}^{\mu}$. The variation of its time component is
\begin{eqnarray}
\delta {\hat A}^a_{0}(x)&=&\int \textrm{d}y \ (\mathcal{D}_0*\epsilon^{2})^b(y)*\{{\hat A}^a_{0}(x),\hat{\pi}^b_0(y)\}\nonumber\\
&=&\int \textrm{d}y \ (\mathcal{D}_0*\epsilon^{2})^b(y)\delta^{ab}*\delta(x-y)\nonumber\\
&=&(\mathcal{D}_0*\epsilon^{2})^{a} \label{A0}
\end{eqnarray}
where we have used the identity (\ref{b1}). The variation of the
space component is found to be $\delta {\hat
A}^a_{i}(x)=(\mathcal{D}_i*\epsilon^{2})^a(x)$. Combining this
with (\ref{A0}) and identifying
$\epsilon^2\rightarrow\hat{\alpha}$ we get (\ref{Amu}). In a
similar way (\ref{si}) and (\ref{sibar}) can be reproduced.

{\it{TLR formalism}}: So far we were discussing about the star
deformed gauge transformation from a general Hamiltonian
formulation which obeys the normal coproduct rule (\ref{tX}). But
as discussed in the previous chapter the action (\ref{lag}) is
also invariant under the undeformed gauge transformations
(\ref{YY1}--\ref{YY}) with the twisted coproduct rule (\ref{co}).

We now present an alternative interpretation of the twisted
coproduct rule (\ref{co}). The results (\ref{gag}) are seen to
follow by using the standard coproduct rule (\ref{tX}) but pushing
the gauge parameter $\hat{\alpha}$ outside the star operation at
the end of the computations \cite{rb,samanta1}. Denoting this manipulation
as,
\begin{eqnarray}
\delta_{\hat{\alpha}}(A*B)\sim(\delta_{\hat{\alpha}}A)*B+A*(\delta_{\hat{\alpha}}B)
\end{eqnarray}
we find
\begin{eqnarray}
\delta_{\hat{\alpha}}({\hat A}_{\mu}*{\hat\psi})&\sim&(\delta_{\hat{\alpha}}{\hat A}_{\mu})*{\hat\psi}+{\hat A}_{\mu}*(\delta_{\hat{\alpha}}{\hat\psi})\nonumber\\
&\sim&(\partial_{\mu}\hat{\alpha}-ig\hat{\alpha}^a[T^a,{\hat A}_{\mu}])*{\hat\psi}+{\hat A}_{\mu}*(-ig\hat{\alpha}^aT^a{\hat\psi})\nonumber\\
&=&\partial_{\mu}\hat{\alpha}{\hat\psi}-ig\hat{\alpha}^a([T^a,{\hat A}_{\mu}]*{\hat\psi})-ig\hat{\alpha}^a({\hat A}_{\mu}*T^a{\hat\psi})\nonumber\\
&=&\partial_{\mu}\hat{\alpha}{\hat\psi}-ig\hat{\alpha}({\hat
A}_{\mu}*{\hat\psi})\label{g.v.}
\end{eqnarray}
which can also be obtained by using the twisted coproduct rule
(\ref{co}).

In order to obtain the undeformed gauge transformations
(\ref{YY1}--\ref{YY}) and the relations (\ref{gag}) we give a
prescription. In the computation of Poisson brackets, the gauge
parameter has to be pushed outside the star operation at the end
of the computations.

 The gauge variation of the time component of ${\hat A}^{\mu}$ field is found by suitably Poisson bracketing with (\ref{generator}) (renaming $\epsilon^{2}$ as $\hat{\alpha}$),
\begin{eqnarray}
\delta_{\hat{\alpha}} {\hat A}^a_{0}(x)&=&\int \textrm{d}y \ (\mathcal{D}_0*\hat{\alpha})^b(y)*\{{\hat A}^a_{0}(x),\hat{\pi}^b_0(y)\}\nonumber\\
&\sim&\int\textrm{d}y \ (\partial_0\hat{\alpha}^{a}-\frac{g}{2}f^{abc}\{{\hat A}_0^b,\hat{\alpha}^{c}\}_*+i\frac{g}{2}d^{abc}[{\hat A}_0^b,\hat{\alpha}^{c}]_*)(y)*\delta(x-y)\nonumber\\
&=&\partial_0\hat{\alpha}^{a}-gf^{abc}{\hat
A}_0^b\hat{\alpha}^{c}\label{tw1}
\end{eqnarray}
where in the last step we put $\hat{\alpha}$ outside the star
product following our prescription. The variation of the space
component is also calculated in a similar way
\begin{eqnarray}
\delta_{\hat{\alpha}} {\hat
A}^a_{i}(x)=\partial_i\hat{\alpha}^{a}-gf^{abc}{\hat
A}_i^b\hat{\alpha}^{c}. \label{tw2}
\end{eqnarray}
Combining (\ref{tw1}) and (\ref{tw2}) we write the gauge variation
in a covariant notation
\begin{eqnarray}
\delta_{\hat{\alpha}} {\hat
A}_{\mu}^a=(\mathcal{D}_{\mu}\hat{\alpha})^a. \label{77}
\end{eqnarray}
The gauge variation of the fermionic field can be obtained in a
similar way. The calculation of the gauge variation of composite
fields needs some care. For example, consider the variation
$\delta_{\hat{\alpha}}({\hat A}_{\mu}*{\hat\psi})$,
\begin{eqnarray}
\delta_{\hat{\alpha}}({\hat A}_{0}(x)*{\hat\psi}(x))&=&T^a\delta_{\hat{\alpha}}({\hat A}_{0}^a(x)*{\hat\psi}(x))\nonumber\\
&\sim&T^a\int \textrm{d}y \ (\mathcal{D}_0*\hat{\alpha}^{a})(y)*\delta(x-y)*{\hat\psi}(x)\nonumber\\
&&-igT^b\int\textrm{d}y \
\hat{\alpha}^{c}(y)*T^c{\hat\psi}(y)*{\hat
A}^b_{0}(x)*\delta(x-y).
\end{eqnarray}
Using the identity (\ref{delta*}) the argument of ${\hat\psi}$ and
${\hat A}^b_{0}$ is changed from $x$ to $y$ to obtain
\begin{eqnarray}
\delta_{\hat{\alpha}}({\hat A}_{0}(x)*{\hat\psi}(x))&\sim&T^a\int \textrm{d}y \ (\mathcal{D}_0*\hat{\alpha}^{a})(y)*{\hat\psi}(y)*\delta(x-y)\nonumber\\
&&-igT^b\int\textrm{d}y \
\hat{\alpha}^{c}(y)T^c*{\hat\psi}(y)*\delta(x-y)*{\hat
A}^b_{0}(y).
\end{eqnarray}
Using the properties (\ref{b1}), (\ref{b2}) and finally removing
the gauge parameter $\hat{\alpha}$ outside the star product we
obtain
\begin{eqnarray}
\delta_{\hat{\alpha}}({\hat
A}_{0}*{\hat\psi})&=&T^a(\partial_0\hat{\alpha}^{a}{\hat\psi}-gf^{abc}\hat{\alpha}^{c}({\hat
A}_{0}^b*{\hat\psi}))-igT^bT^c\hat{\alpha}^{c}({\hat
A}_{0}^b*{\hat\psi}).
\end{eqnarray}
Using the symmetry algebra (\ref{dabc},\ref{trace}) it is easy to
check that above result is the time component of the equation
(\ref{gag}). similar result can also be found for the space part.
The gauge variations of the other composites are computed in the
same way reproducing the results (\ref{gag}) obtained by using the
twisted coproduct rule.

\section{Noncommutative Gravity and Black Hole Physics}
The renewed interest in NC spacetime is mainly due to its
relevance in quantum gravity research. Formulation of gravity
theories over NC spacetime thus attracted a huge attention in the
literature \cite{szabog, nicolini}. There are various attempts to
fit General theory of Relativity (GTR) in the context of NC space
time. In \cite{Chamseddine:2000si} for example  a  deformation of
Einstein's gravity was studied using a  construction  based on
gauging the noncommutative SO(4,1) de Sitter group and the SW map
\cite{sw} with subsequent contraction to ISO(3,1). Another
construction of a noncommutative gravitational theory was proposed
in \cite{Aschieri:2005yw}. Recently noncommutative gravity
has been connected with stringy perspective\cite{ag}.In all these
works the leading order noncommutative effects appear in the
second order in the NC parameter $\theta$.

    The introduction of non-commutativity spoils the symmetry under general
    coordinate transformation. However, the NC geometry is compatible with a
    restricted class of coordinate transformations which is volume preserving.
    The corresponding formulation of NC gravity \cite{CK1} , often referred
    to as the minimal theory, brings us to the realm of Unimodular
    Gravity \cite{UNI}. Initially, the leading order correction was
    reported to be linear in $\theta$ in this work. The model was
    reconsidered in \cite{MS} where it was shown by explicit
    construction that the first order correction actually vanishes.
    The second order corrections were later computed \cite{CK2}.
    The noncommutative structure in \cite{CK1, MS, CK2} is
    constant $\theta_{\mu\nu}$. Subsequently, NC gravity in
    the minimal theory approach was developed with
    $\theta_{\mu\nu}$ having a Lie-algebraic structure
    \cite{BMS}. The vanishing of the first order correction
    was again observed. The same phenomenon is observed in
in calculations from various angles \cite {HR,K}. Below we will
show how the first order term vanishes by reviewing results from
\cite{BMS} in the Lie algebraic form of NC spacetime. The results
for canonical noncommutativity \cite{MS} may be obtained from
these by a limiting procedure.

          The methodology of direct generalisation of gravity to NC
          spacetime that can yield results useful for different
          phenomenology is based on a perturbative expansion in
          the NC parameter $\theta$. The leading order NC correction
          being second order in the NC parameter is indeed small.
          However, this has been shown to have important
          phenomenological consequences \cite {CTZ} where NC
          generalization of the Schwarzschild solution has
          been worked out. Later on the impact of the NC effect
          on Charged black hole has been analysed \cite{MS1, CTZ1}.
          Below we will review the results of \cite{MS1} where the
          deformed Reissner-Nordstrom solutions are given. The results
          for the Schwarzschild solutions given in \cite{CTZ} are
          obtainable by a limiting procedure from these results.
          Considering the complex steps involved in the computation
          of the NC corrections this correspondence is indeed notable.

      The perturbative expansion employed in the NC gravity theories
      involve a cut off which removes the original nonlocality in the
      NC theories. This perturbative expansion is however an essential
      feature of incorporating NC effects in gravity. In the last few years an interesting study has been done in the literature\cite{hsyang,ban-yang} where gravity is interpreted in a completely different manner within a noncommutative framework. Here gravity is not incorporated in the theory by hand rather it turns out to be an emergent structure from the electromagnetic phenomena in noncommutative spacetime. An alternative
      procedure is to consider the effect of the fuzziness in the
      level of the mean values \cite{nicolini,nicrev,grezia}. This "noncommutativity
      inspired" methodology has been pursued in \cite{B}. We will
      conclude our review of NC gravity with a few results obtained in this approach.

\subsection{Lie Algebraic Noncommutative Gravity} We consider noncommutativity of the form
\begin{eqnarray}
[\hat{x}^{\mu},\hat{x}^{\nu}]&=&i\theta^{\mu\nu}(\hat{x})\nonumber\\
&=&i\theta f^{\mu\nu}_{ \ \ \ \lambda}\hat{x}^{\lambda}
\label{lie}
\end{eqnarray}
where $f^{\mu\nu}_{ \ \ \ \lambda}$ are the structure constants.
For consistency these constants assume a Lie algebraic structure
so that $f_{\mu\nu\lambda}$ is antisymmetric in all the three
indices.

The formulation of gravity on NC space time poses problems. This
is seen by considering the general coordinate transformation,
\begin{eqnarray}
\hat{x}^{\mu}\rightarrow
\hat{x}'^{\mu}=\hat{x}^{\mu}+\hat{\xi}^{\mu}(\hat{x}) \label{xxi}
\end{eqnarray}
and realising that, for arbitrary $\hat{\xi}^{\mu}(\hat{x})$, it
is not compatible with the algebra (\ref{lie}). However, as in the
canonical case\cite{CK1}, it is possible to find a restricted
class of coordinate transformations (\ref{xxi}) which preserves
the Lie -- algebraic noncommutative algebra. We  exploit the Weyl
-- Wigner correspondence \cite{szabo} to work in the deformed
phase space with the ordinary multiplication substituted by the
corresponding star product.

    The noncommutative coordinates $\hat{x}^{\mu}$ satisfying (\ref{lie})
are the generators of an associative algebra ${\cal{A}}_x$.
According to the Weyl correspondence we can associate an element
of ${\cal{A}}_x$ with a function $f(x)$ of classical variables
$x^{\mu}$ by the unique prescription. The *-product between two
classical functions $f(x)$ and $g(x)$ is denoted by $f*g$ and is
defined by the requirement
 functions with their product defined by the star product.
When the generators satisfy the Lie structure the star product is
explicitly given by \cite{madore,Jurco:2000ja}
 \begin{eqnarray}
f(x)*g(x)=e^{\frac{i}{2}x^{\lambda}g_{\lambda}(i\frac{\partial}{\partial
x'},i\frac{\partial}{\partial
x''})}f(x')g(x'')|_{(x',x'')\rightarrow x} \label{star}
\end{eqnarray}
where $g_{\lambda}$ is defined by,
\begin{eqnarray}
e^{ik_{\lambda}\hat{x}^{\lambda}}e^{ip_{\lambda}\hat{x}^{\lambda}}=
e^{i\{k_{\lambda}+p_{\lambda}+\frac{1}{2}g_{\lambda}(k,p)\}\hat{x}^{\lambda}}
\end{eqnarray}
The explicit form of $g_{\lambda}(k,p)$ is obtained as
\begin{eqnarray}
g_{\lambda}(k,p)&=&-\theta k_{\mu}p_{\nu}f^{\mu\nu}_{ \ \ \
\lambda}+\frac{1}{6}\theta^2k_{\mu}p_{\nu}(p_{\sigma}-k_{\sigma})f^{\mu\nu}_{
\ \ \ \delta}f^{\delta\sigma}_{ \ \ \
\lambda}\nonumber\\&&+\frac{1}{24}\theta^3(p_{\sigma}k_{\beta}+
k_{\sigma}p_{\beta})k_{\mu}p_{\nu}f^{\mu\nu}_{ \ \ \
\delta}f^{\delta\sigma}_{ \ \ \ \alpha}f^{\alpha\beta}_{ \ \ \
\lambda}+... \label{95}
\end{eqnarray}

Now in order to preserve the noncommutative algebra (\ref{lie}) under the general coordinate transformation (\ref{xxi}), $\xi^{\mu}$ must satisfy
the condition,
\begin{eqnarray}
\hat{\xi}^{\mu}(x)=f^{\mu\alpha}_{ \ \ \
\beta}x^{\beta}\partial_{\alpha}g(x) \label{restrtrans}
\end{eqnarray}
so that
\begin{eqnarray}
[\hat{x}'^{\mu},\hat{x}'^{\nu}]&=&i\theta f^{\mu\nu}_{ \ \ \ \lambda}\hat{x}'^{\lambda}
\label{llie}
\end{eqnarray}
 Here the symbol $g(x)$ which appears in (\ref{restrtrans}) denotes an arbitrary function. From (\ref{restrtrans}) we find that
\begin{eqnarray}
\partial_{\mu}\hat{\xi}^{\mu}(x) = 0,\nonumber
\end{eqnarray}
which implies that the Jacobian of the transformations (\ref{xxi}) is then unity. In
other words the transformations are volume preserving. The
corresponding theory thus belongs to the noncommutative version of
unimodular gravity.

We have now all the tools at our disposal to develop the
commutative equivalent theory of noncommutative gravity in the
framework of
Poincare gauge theory of  gravity \cite{szabog}. 
The corresponding
 noncommutative gauge transformation can be decomposed in the following way
\begin{eqnarray}
\hat{\Lambda}(\hat{x})=\hat{\xi}^{\mu}(\hat{x})p_{\mu}+\frac{1}{2}
\hat{\lambda}^{ab}(\hat{x})\Sigma_{ab}. \label{gauge}
\end{eqnarray}
Here $\hat{\xi}^{\mu}$ is the local translation of the tetrad
which must be restricted to the form given in equation
(\ref{restrtrans}) in order to preserve the noncommutative algebra
(\ref{lie}). The parameters $\hat{\lambda}^{ab}(\hat{x})$
characterize the local Lorentz transformations at $\hat{x}$ with
$\Sigma_{ab}$
as the generators of the Lorentz group.  In what follows
we will assume the vector representation of $Sigma_{ab}$.
As is usual we will denote the general coordinates by the Greek
indices and components with respect to the tetrad by Latin
indices. Corresponding to the noncommutative gauge transformations
(\ref{gauge}) we introduce the gauge potential
\begin{eqnarray}
\hat{A}_{a}(\hat{x})=(\hat{D}_a)=i\hat{E}^{\mu}_a(\hat{x})p_{\mu}+\frac{i}{2}\hat{\omega}_a^{
\ bc}(\hat{x})\Sigma_{bc}\label{gp}
\end{eqnarray}
where $E^{\mu}_{a}(\hat{x})$ are the components of the
noncommutative tetrad $\hat{E}_a$ which are also the gauge fields
corresponding to general coordinate transformations and
$\hat{\omega}_a^{ \ bc}(\hat{x})$ are the spin connection fields
associated with local Lorentz invariance. Since $p_{\mu} =
-i\partial_{\mu}$, the noncommutative tetrad maps trivially on the
commutative one \cite{CK1}. Assuming the gauge transformations and
the spin connection fields in the enveloping algebra approach
\cite {Jurco:2000ja} we can write the order $\theta$ corrections
to these potentials. Using these, the field strength tensor is
worked out. This is identified with the NC Riemann tensor under
zero torsion. From the NC Riemann tensor the NC Ricci Scalar is
straightforwardly computed.
Thus we can write
\begin{eqnarray}
\hat{R}_{ab}=R_{ab}+R^{(1)}_{ab}+\mathcal{O}(\theta^2)
\end{eqnarray}
where the correction term is obtained as,
\begin{eqnarray}
R^{(1)}_{ab}&=&\frac{1}{2}\theta^{cd}\{R_{ac},R_{bd}\}-\frac{1}{4}
\theta^{cd}\{\omega_c,(\partial_d+\mathcal{D}_d)R_{ab}\}\nonumber\\&&
+\frac{1}{2}\theta_{bc}\theta^{de}\partial_e\theta^{fc}\{R_{af},
\omega_d\}-\frac{1}{2}\theta_{ac}\theta^{de}\partial_e\theta^{fc}\{R_{bf},\omega_d\}.
\label{67}
\end{eqnarray}
The Ricci tensor $\hat{R}_{a}^{ \ c}=\hat{R}_{ab}^{ \  \ bc}$ and
the Ricci scalar $\hat{R}=\hat{R}_{ab}^{ \  \  ab}$ are formed to
construct the action
\begin{eqnarray}
S&=&\int d^4x \ \frac{1}{2\kappa^2}\hat{R}(\hat{x})\\
&=&\int d^4x \
\frac{1}{2\kappa^2}\left(R(x)+R^{(1)}(x)\right)+\mathcal{O}(\theta^2).
\end{eqnarray}
The first order correction term to the Lagrangian is
\begin{eqnarray}
R^{(1)}(x)=R^{(1)ab}_{ab}=[R^{(1)}_{ab}]^{ab}
\end{eqnarray}
It is convenient to arrange the correction as
\begin{eqnarray}
[R^{(1)}_{ab}]^{ab}=\mathcal{R}_1+\mathcal{R}_2+\mathcal{R}_3+\mathcal{R}_4.
\label{correction}
\end{eqnarray}
where $\mathcal{R}_1,...,\mathcal{R}_4$ correspond to the
contributions coming from the four pieces appearing on the right hand side
of (\ref{67}) in the same order.
Exploiting the various symmetries of the Riemann Tensor, spin
connection and the noncommutative structure $\theta^{ab}$ we can
easily show that both $\mathcal{R}_1$ and $\mathcal{R}_2$
individually vanish. Note that these terms do not depend on the
coordinate dependence of $\theta^{ab}$ and will remain valid for
canonical noncommutative structure. This is the canonical limit of
\cite{CK1,MS}. The last two terms on the r.h.s. of
(\ref{correction})
 owe their existence to the Lie -- algebraic noncommutativity
assumed in the present work. Most significantly
\begin{eqnarray}
\mathcal{R}_3+\mathcal{R}_4=0
\end{eqnarray}
The vanishing of first order correction of the Ricci scalar for constant noncommutativity is well known \cite{MS}. Here we  find the same result for
 the Lie-algebraic structure.
 From the present analysis it is clear
that various symmetries of the Riemann tensor and the spin connection
of the commutative theory are responsible for the non existence of order $\theta$ correction
. Thus it seems that the zero value of 
the first order correction is due to the underlying symmetries of
space time which will presumably hold for more general
noncommutative structure. However, we are unable to give a
definitive proof of this.

\subsection{Noncommutativity Inspired Black Hole Physics}
We will now study some applications of NC gravity in Black Hole
physics.  We have seen that a direct generalisation of Einstein's
theory of gravity to noncommutative spacetime results in NC
correction which is a second order effect. However, this small
correction incorporated in different phenomenology has been shown
to produce important physical effects \cite{CTZ,MS1,CTZ1,stern}.
In these works the space-time of noncommutative theory is taken to
be of Minkowski type, endowed with spherical noncommutative
coordinates. A deformation of the gravitational field is
constructed by gauging the noncommutative de Sitter $SO(4,1)$
group \cite{Chamseddine:2000si} and using Seiberg-Witten (SW) map
\cite{sw}. The deformed gravitational gauge potentials (tetrad
fields) $\hat{e} _{\mu }^{a}\left( {x,\theta }\right) $ are
obtained by contraction of the noncommutative gauge group
$SO(4,1)$ to the Poincar\'{e} (inhomogeneous Lorentz) group
$ISO(3,1)$. The fields are expanded in perturbative series where
the different terms of the series are obtained from the
commutative solution of the metric. The deformed gauge fields  up
to the second order in the noncommutativity parameters $\theta
^{\mu \nu}$ are found. The correction terms require the
commutative tetrad fields of the de Sitter gauge theory of
gravitation over Minkowski spacetime. These solutions are found
using a spherically symmetric ansatz and solving the corresponding
Einstein equations. From the NC tetrad fields $\hat{e} _{\mu
}^{a}\left( {x,\theta }\right) $ we construct the NC
Reissner--Nordstrom metric $\hat{g}_{\mu \nu } \left( {x,\theta
}\right) $. Naturally, the nontrivial correction starts from the
second order. We explicitly calculate this leading NC correction
term to the Reissner--Nordstrom metric. These solutions have been
used to study the NC deformation of the charged black hole
solutions \cite{MS1,CTZ1}. The deformed metric is calculated by
the formula:
\begin{equation}
\hat{g}_{\mu \nu }\left( {x,\theta }\right) =\frac{1}{2}\,\eta
_{a\,b}\,\left( {\hat{e}_{\mu }^{a}\star \hat{e}_{\nu
}^{b}{}^{+}+\hat{e} _{\nu }^{b}\star \hat{e}_{\mu
}^{a}{}^{+}}\right) .  \label{3.14}
\end{equation}

Here $+$ means complex conjugation. Note that Chamseddine's formalism involves extra complex degrees of freedom. These are necessary to close the gauge algebra in the noncommutative framework. One is then faced with the problem of the spurious degrees of freedom that remain after the commutative limit is taken. For a discussion of this point see \cite{Chamseddine:2000si,chams}.

The constant antisymmetric matrix $\theta^{\mu \nu}$ can always be
rotated to a skew-diagonal form \cite{szabog}. We further assume
vanishing noncommutativity in the time-space sector, which is
quite usual in
the literature. 
The non-zero components of the tetrad fields $\hat{e}_{\mu}^
{a}\left( {x,\theta }\right) $ corresponding to this NC structure
can be easily worked out using GRTensor II package of Maple. Then
using the definition of the metric (\ref{3.14}) we arrive at the
following non-zero components of the deformed metric
$\hat{g}{}_{\mu \nu}$ up to the second order in $\theta$.
The explicit form of the non-zero components of the NC Reissner--Nordstrom
 metric computed in this way \cite{MS1} are
\begin{eqnarray}
\hat{g}_{00}&=& -\left( {1-\frac{2M }{r} +
\frac{Q^{2}}{r^{2}}}\right) - \frac{1}{r^{6}}\left[Mr^{3} -
\frac{11 M^{2} + 9 Q^{2}}{4}r^{2} - \frac{17 M Q^{2}}{4} r -
\frac{7 Q^{4}}{2}\right]\theta^{2}+ O(\theta^{4})
\nonumber \\
\hat{g}_{11}&=& \frac{1}{\left( {1-\frac{2M }{r} + \frac{Q^{2}}{r^{2}}}\right)} + \frac{\left[- 2 M r^{3} + 3\left(M^{2} + Q^{2}\right) r^{2} - 6 M Q^{2} r + 2 Q^{4} \right]}{4 r^{2} \left(r^{2} - 2 M r + Q^{2}\right)^{2}} \theta^{2} + O(\theta^{4})\nonumber\\
\hat{g}_{22}&=& r^{2} + \frac{1}{16}\left[1 - \frac{15 M}{r} + \frac{26 Q^{2}}{r^{2}} + \frac{4 \left(M r - Q^{2}\right)^{2}}{r^{2} \left(r^{2} - 2 M r + Q^{2}\right)}\right]\theta ^{2} +O(\theta ^{4}) \nonumber \\
\hat{g}_{33}&=& r^{2} \sin^{2}\phi+ \frac{1}{16}\left[\frac{4 r^{2} \left(M^{2} - M r\right) + 8 Q^{2} \left( r^{2} - 2 M r\right) + 8 Q^{4} }{r^{2} \left(r^{2} - 2 M r + Q^{2}\right)^{2}} \sin^{2} \phi + \cos^{2} \phi\right]\theta ^{2}+ O(\theta ^{4}) \nonumber \\
\label{4.5}
\end{eqnarray}
If we substitute $Q = 0$ in our expressions (\ref{4.5}) the
solutions exactly reduces to the NC Schwarzschild solutions
\cite{CTZ}.

 The NC corrections to the charged black hole solution may be obtained from the deformed metric.
 For instance we consider the event horizon. In commutative space-time we can identify the event
 horizons by following radial null curves and locating the radius at which $\frac{dt}{dr}$ becomes
 infinity. Following this and remembering that our event horizons should go to the commutative results in the
 limit $\theta \to 0$ we define the event horizons of the NC R--N metric from $g_{00} = 0$.
From our NC R--N solutions (\ref{4.5}) it is straightforward to
derive
\begin{eqnarray}
r^{2} - 2 mr + Q^{2} = -\frac{\theta^{2}}{4 r^{4}}\left(4 m r^{3}
- 11 m^{2} r^{2} + 9 Q^{2} r^{2} - 17 m Q^{2} r + 14 Q^{4}\right)
\label{H1}
\end{eqnarray}
the solutions to which give the horizon radii. Naturally we look
for solutions correct up to second order in $\theta$. The required
solutions are
\begin{eqnarray}
r_{+} &=& M + \sqrt{M^{2} - Q^{2}} + \frac{\theta^{2}}{2}\frac{A_{+}}{\sqrt{M^{2} - Q^{2}}} \nonumber\\
r_{-} &=& M - \sqrt{M^{2} - Q^{2}} -
\frac{\theta^{2}}{2}\frac{A_{-}}{\sqrt{M^{2} - Q^{2}}} \label{H2}
\end{eqnarray}
with $A_{+}$ and $A_{-}$ given by
\begin{eqnarray}
A_{+} &=& \frac{6 M^4 + 10 M^3 \left( M^2 - Q^2 \right)^{(1/2)} +
36 Q^2 M^2
 - 4 M \left( M^2 - Q^2 \right)^{(3/2)}}{4 \left(M + \sqrt{M^2 - Q^2}\right)^4}\nonumber\\
&& + \frac{35 Q^2 M \left( M^2 - Q^2 \right)^{(1/2)} + 5 Q^4}{4
\left(M + \sqrt{M^2 - Q^2}\right)^4}
 \nonumber\\
A_{-} &=& \frac{6 M^4 - 10 M^3 \left( M^2 - Q^2 \right)^{(1/2)} +
36 Q^2 M^2
 + 4 M \left( M^2 - Q^2 \right)^{(3/2)}}{4 \left(M - \sqrt{M^2 - Q^2}\right)^4}\nonumber\\
&& - \frac{35 Q^2 M \left( M^2 - Q^2 \right)^{(1/2)} + 5 Q^4}{4
\left(M - \sqrt{M^2 - Q^2}\right)^4} \label{H3}
\end{eqnarray}
Note that these solutions properly map to the familiar
(commutative) R--N horizon
\begin{eqnarray}
r_{\pm} = M \pm \sqrt{M^{2} - Q^{2}} \label{H4}
\end{eqnarray}
in the limit $\theta \to 0$. As a result of the NC effect the
distance between the event horizon radii increases. Also similar
calculations have been performed for the Robertson Walker metric
\cite{stern}.

In the above we have discussed the deformation of general
relativity in noncommutative space time. Our approach was
perturbative (through the SW map\cite{sw}). Thus the essential
nonlocality inherent in the NC field theories was not taken into
account. Moreover in this approach it is not clear how to
incorporate the basic symmetries in the model. As mentioned
earlier an alternative approach was proposed where the effect of
noncommutativity has been introduced in the level of the mean
value\cite{nicolini,cite3,cite4}. In the following section some thermodynamic
applications of this approach is reviewed.

\subsection{Coherent State Based Approach to the Noncommutative Black Hole Physics}
The alternative approach towards incorporating NC effects is based
on the coherent state formalism\cite{glauber} of quantum optics.
The idea is to modify the volume density to introduce the
fuzziness corresponding to the NC structure. The method is
exemplified in the following.

{\it{NC Schwarzschild black hole}}: A point particle of mass $M$
is described by the following volume density
\begin{eqnarray}
\rho=M\delta^3(\bf r) \label{msdi}
\end{eqnarray}
but in a noncommutative space, concept of point does not exist.
Instead, there is a smearing which comes as a consequence of
position-position uncertainty relation. We introduce the
noncommutative correction in (\ref{msdi}) by replacing the Dirac
delta function by a Gaussian distribution of minimal width
$\sqrt{\theta}$
\begin{eqnarray}
\rho_\theta=\frac{M}{(4\pi\theta)^{\frac{3}{2}}}e^{-\frac{r^2}{4\theta}}.
\label{2.222}
\end{eqnarray}
where the noncommutative parameter $\theta$ is considered to be a
small ($\sim{\textrm{Planck length}}^2$) positive number. Using
the above expression one can find the mass of a black hole by
integrating (\ref{2.222}) over a volume of radius $r$. This is
found to be,
\begin{eqnarray}
m_\theta(r)=\int_0^r 4\pi r'^2 \rho_{\theta}(r')dr' =
\frac{2M}{\sqrt{\pi}}\gamma(\frac{3}{2},\frac{r^2}{4\theta})
\label{2.444}
\end{eqnarray}
where $\gamma(\frac{3}{2},\frac{r^2}{4\theta})$ is the lower
incomplete gamma function defined in the following way
\begin{eqnarray}
\gamma(a,x)=\int_0^x t^{a-1}e^{-t}dt
\end{eqnarray}
In the $\theta\rightarrow 0$ limit it becomes the usual gamma
function
($\Gamma_{\textrm{total}}(a)=\gamma(a,x)+\Gamma(a,x)$){\footnote{$\Gamma(a,x)=\int_x^{\infty}
t^{a-1}e^{-t}dt$ is the upper incomplete gamma function}.}
\begin{eqnarray}
\Gamma_{\textrm{total}}=\int_0^{\infty} t^{a-1}e^{-t}dt
\end{eqnarray}
and $m_\theta(r)$ of (\ref{2.444}) reduces to $M$. Substituting
(\ref{2.444}) in the mass term of the Schwarzschild space time
\begin{eqnarray}
ds^2 = -(1-\frac{2M}{r})dt^2 + (1-\frac{2M}{r})^{-1}dr^2 + r^2
d\Omega^2. \label{2.1}
\end{eqnarray}
we get the noncommutative Schwarzschild metric,
\begin{eqnarray}
ds^2 =
-\Big(1-\frac{4M}{r\sqrt{\pi}}\gamma(\frac{3}{2},\frac{r^2}{4\theta})\Big)dt^2
+
\Big(1-\frac{4M}{r\sqrt{\pi}}\gamma(\frac{3}{2},\frac{r^2}{4\theta})\Big)^{-1}dr^2
+ r^2 d\Omega^2 \label{2.555}
\end{eqnarray}
Above metric can also be interpreted as the solution of the
Einstein equation
$(G_{\theta})^{\mu\nu}=8\pi(T_{\theta})^{\mu\nu}$. Here the
$T_{\theta}$ is the energy momentum tensor
\begin{eqnarray}
(T_{\theta})_{\mu}^{\nu}={\textrm{diag}}[-\rho_{\theta},p_r,p',p']
\end{eqnarray}
where, $p_r=-\rho_{\theta}$ and
$p'=p_r-\frac{r}{2}\partial_r\rho_{\theta}$ so that
$(T_{\theta})_{\mu ; \nu}^{\nu}=0$. The expression of $p'$
\begin{eqnarray}
p'=\left(\frac{r^2}{4\theta}-1\right)\frac{M}{(4\pi\theta)^{\frac{3}{2}}}\textrm{e}^{-\frac{r^2}{4\theta}}
\end{eqnarray}
reduces to $p_r$ when $r\ll \sqrt{\theta}$. In the other limit
($r\gg \sqrt{\theta}$) the energy momentum tensor is zero and one
recovers the Schwarzschild vacuum. Thus in the small length scale
the radial pressure nullifies the gravitational attraction and
hence collapse is prevented. Previously such phenomenon was
associated with the presence of a de-Sitter metric inside the
black hole\cite{fro,dym}. Here in the small $r$ limit
\begin{eqnarray}
-g_{tt}=1-\frac{Mr^2}{3\sqrt{\pi}\theta^{\frac{3}{2}}}\label{deS}
\end{eqnarray}
For (\ref{deS}) the metric (\ref{2.555}) becomes a de-Sitter
metric with cosmological constant
\begin{eqnarray}
\Lambda=\frac{M}{3\sqrt{\pi}\theta^{\frac{3}{2}}}
\end{eqnarray}
having the following constant curvature
\begin{eqnarray}
R=\frac{4M}{\sqrt{\pi}\theta^{\frac{3}{2}}}
\end{eqnarray}
As a result we get a de-Sitter core of positive curvature
surrounding the singularity.

Note that the metric (\ref{2.555}) is stationary, static and
spherically symmetric in nature -- this fact has immense
importance in the subsequent thermodynamic analysis. The event
horizon ($r_h$) for the metric (\ref{2.555}) can be found by
setting $g_{tt}(r_h)=0$. The result obtained is
\begin{eqnarray}
r_h =
\frac{4M}{\sqrt{\pi}}\gamma\Big(\frac{3}{2},\frac{r_h^2}{4\theta}\Big)
\label{2.6}
\end{eqnarray}
which can not be solved in a closed form.
\begin{figure}[i]
\centering
\includegraphics[angle=0,width=15cm,height=15cm]{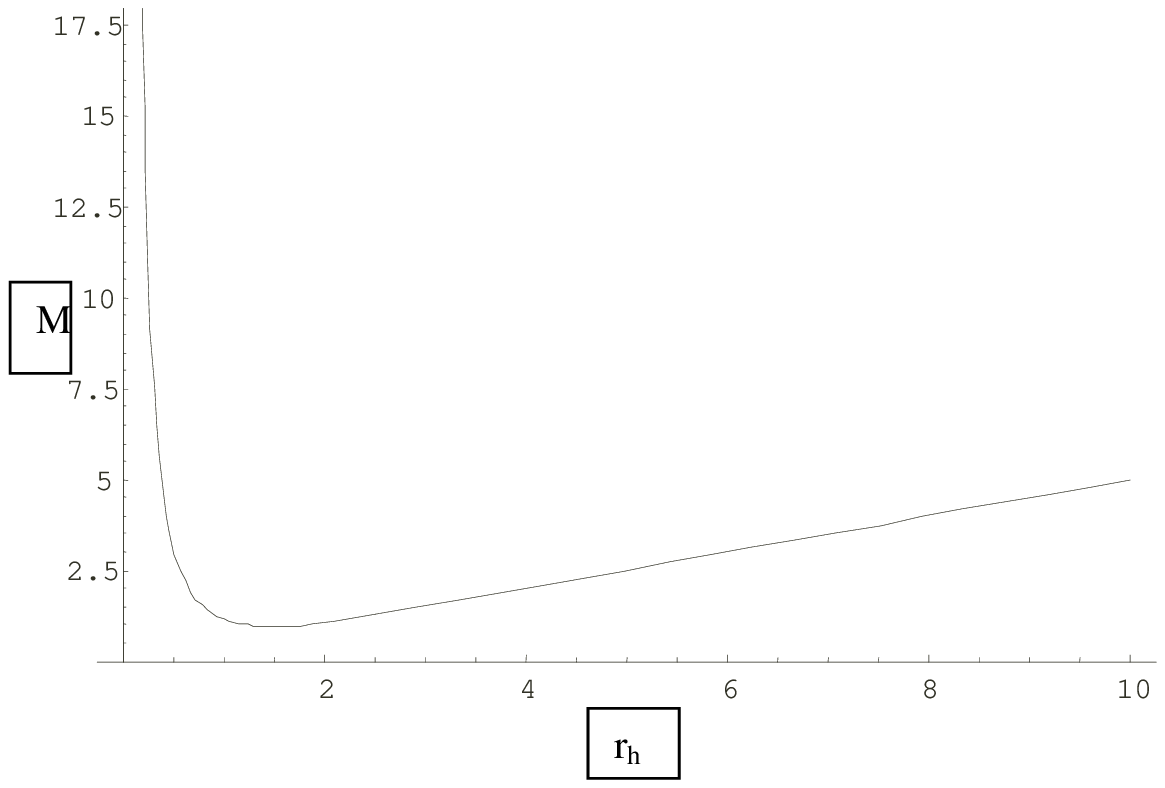}
\caption[]{{\it{$M$ Vs. $r_h$ plot.\\
$M$ is plotted in units of $2\sqrt\theta$ and $r_h$ is plotted in units of $2\sqrt\theta$\\
For eq. (\ref{2.6}).}}} \label{fig1}
\end{figure}

       In Figure (\ref{fig1}) we plot the black hole mass $M$ as a function of $r_h$
       (for equation (\ref{2.6})). Clearly at $r_h=3.0\sqrt\theta$, $M$ is minimum
       ($M_{{\textrm{min}}}=1.9\sqrt\theta$) and when $M<M_{{\textrm{min}}}$ no
       horizon is possible. This counter intuitive result has no analogy in the
       standard commutative space description of Schwarzschild black hole
       \cite{nicolini,kim,BB}. When $M>M_{{\textrm{min}}}$ two different
       horizons appear one of them is outer (event) horizon and the
       other is inner (Cauchy) horizon. These two horizons merge as
       $M\rightarrow M_{{\textrm{min}}}$. In the special case when
       $M$ is very large compared to $M_{{\textrm{min}}}$ the inner
       horizon shrinks to zero and the outer horizon approaches the
       Schwarzschild radius $2M$. Detailed discussions on these
       results may be found in \cite{nicolini,kim}.

In the large radius regime ($\frac{r_h^2}{4\theta}>>1$) the
asymptotic expansion of the lower incomplete $\gamma$ function
\begin{eqnarray}
\gamma(\frac{3}{2},x)&=&\Gamma_{{\textrm{total}}}(\frac{3}{2})-\Gamma(\frac{3}{2},x)
\nonumber
\\
&\simeq&\frac{\sqrt{\pi}}{2}\Big[1-e^{-x}\sum_{p=0}^{\infty}\frac{x^{\frac{1-2p}{2}}}
{\Gamma_{{\textrm{total}}}(\frac{3}{2}-p)}\Big] \label{app4}
\end{eqnarray}
is used to solve (\ref{2.6}) by iteration. Keeping up to the order
$\frac{1}{\sqrt{\theta}}e^{-\frac{M^2}{\theta}}$, we find
\begin{eqnarray}
r_h\simeq2M\Big(1-\frac{2M}{\sqrt{\pi\theta}}e^{-\frac{M^2}{\theta}}\Big).
\label{2.7}
\end{eqnarray}

{\it{Black hole temperature}}: The surface gravity $(\kappa(r_h))$
for the classical noncommutative Schwarzschild spacetime
(\ref{2.555}) is given by

\begin{eqnarray}
\kappa(r_h) =\frac{1}{2}[\frac{dg_{00}}{dr}]_{r=r_h}=
\frac{1}{2}\Big[\frac{1}{r_h}-\frac{r^2_h}{4\theta^{\frac{3}{2}}}
\frac{e^{-\frac{r^2_h}{4\theta}}}{\gamma\Big(\frac{3}{2},\frac{r^2_h}{4\theta}\Big)}\Big].
\label{4.3}
\end{eqnarray}

The expression of surface gravity (\ref{4.3}) is true when there
is no back reaction. In order to find the appropriate modification
in the presence of back reaction we first fix the units
$G=c=k_{\textrm B}=1$, in which{\footnote {Planck  length  $l_P
=(\hbar G/c^3)^{1/2}$, Planck  mass  $M_P =(\hbar c/G)^{1/2}$}}
Planck length $l_{\textrm p}$=Planck mass $M_{\textrm p}$
$=\sqrt{\hbar}$. Since a loop expansion is equivalent to an
expansion in powers of the Planck constant, the one loop back
reaction effect in the surface gravity is written as,
\begin{eqnarray}
{\cal{K}}(r_h) = \kappa(r_h)+\xi\kappa(r_h) \label{xi}
\end{eqnarray}
where $\xi$ is a dimensionless constant having magnitude of the
order of ${\hbar}$. From dimensional arguments, therefore, it has
the structure,
\begin{eqnarray}
\xi=\beta\frac{M_{\textrm p}^2}{m^2_\theta}
\end{eqnarray}
where $\beta$ is a pure numerical factor. In the commutative
picture $\beta$ is known to be related to the trace anomaly
coefficient\cite{Lousto,Fursaev}. Putting this form of $\xi$ in
(\ref{xi}) we get,
\begin{eqnarray}
{\cal{K}} = \kappa(r_h)\Big(1+\beta\frac{M_{\textrm
p}^2}{m^2_\theta}\Big) \label{4.1z}
\end{eqnarray}
A similar expression was obtained earlier in \cite{York,Lousto}
for the commutative case. Eq. (\ref{4.1z}) is recast as,
\begin{eqnarray}
{\cal{K}} = \kappa(r_h)\Big(1+\frac{\alpha}{m^2_\theta(r_h)}\Big)
\label{4.1}
\end{eqnarray}
where $\alpha=\beta M_{\textrm p}^2$. 

   In order to write the above equation completely in terms of $r_h$ we have
   to express the mass $m_\theta$ in terms of $r_h$. For that we compare eqs.
   (\ref{2.444}) and (\ref{2.6}) to get,
\begin{eqnarray}
m_{\theta}(r_h)=\frac{r_h}{2} \label{4.4}
\end{eqnarray}
It is noteworthy that the structure of the above equation is
identical to its commutative version.

Substituting (\ref{4.4}) in (\ref{4.1}) we get the value of
modified noncommutative surface gravity
\begin{eqnarray}
{\cal{K}} =
\frac{1}{2}\Big[\frac{1}{r_h}-\frac{r^2_h}{4\theta^{\frac{3}{2}}}
\frac{e^{-\frac{r^2_h}{4\theta}}}{\gamma\Big(\frac{3}{2},\frac{r^2_h}
{4\theta}\Big)}\Big]\Big(1+\frac{4\alpha}{r^2_h}\Big) \label{04.5}
\end{eqnarray}
So the noncommutative Hawking temperature including the effect of
back reaction is given by,
\begin{eqnarray}
T_h=\frac{{\cal{K}}}{2\pi}=\frac{1}{4\pi}\Big[\frac{1}{r_h}-\frac{r^2_h}
{4\theta^{\frac{3}{2}}}
\frac{e^{-\frac{r^2_h}{4\theta}}}{\gamma\Big(\frac{3}{2},\frac{r^2_h}
{4\theta}\Big)}\Big]\Big(1+\frac{4\alpha}{r^2_h}\Big) \label{4.66}
\end{eqnarray}
If the back reaction is ignored (i. e. $\alpha =0$), the
expression for the Hawking temperature is
\begin{eqnarray}
T_h=\frac{1}{4\pi}\Big[\frac{1}{r_h}-\frac{r^2_h}{4\theta^{\frac{3}{2}}}
\frac{e^{-\frac{r^2_h}{4\theta}}}{\gamma\Big(\frac{3}{2},\frac{r^2_h}{4\theta}\Big)}\Big]
\label{1.08}
\end{eqnarray}
which agrees with the temperature obtained in \cite{nicolini}.
Also for the commutative space limit ($\theta\rightarrow 0$) we
get
\begin{eqnarray}
T_h=\frac{1}{4\pi r_h}(1+\frac{4\alpha}{r_h^2})
\end{eqnarray}
Since in the limit $\theta\rightarrow 0$, $r_h\rightarrow 2M$
above equation is written in the standard form
\begin{eqnarray}
T_h=T_H(1+\frac{\alpha}{M^2})\label{4.61}
\end{eqnarray}
where $T_H=\frac{1}{8\pi M}$ is the semiclassical temperature of
the Schwarzschild black hole. This matches with the result
previously found in \cite{Fursaev,majhi}.

\begin{figure}[t]
\centering
\includegraphics[angle=0,width=15cm,height=15cm]{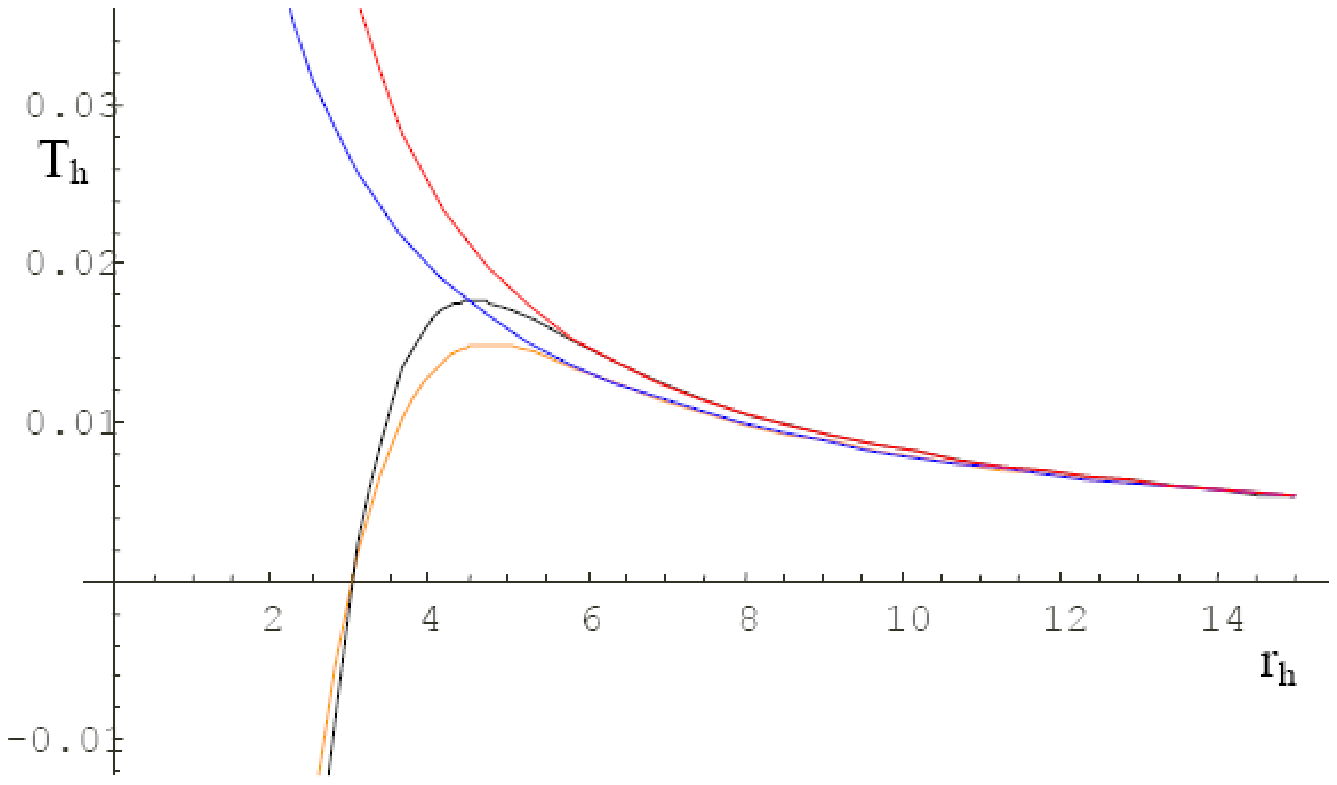}
\caption[]{\it{{$T_h$ Vs. $r_h$ plot (Here $\alpha=\theta$,
$\alpha$ and
$\theta$ are positive).\\
$r_h$ is plotted in units of $\sqrt\theta$ and $T_h$ is plotted in
units
of $\frac{1}{\sqrt\theta}$.\\
Red curve: $\alpha\neq 0, \theta=0$.\\
Blue curve: $\alpha= 0, \theta= 0$.\\
Black curve: $\alpha\neq 0, \theta\neq 0$.\\
Yellow curve: $\alpha= 0, \theta\neq 0$.}}} \label{fig2}
\end{figure}

\begin{figure}[t]
\centering
\includegraphics[angle=0,width=15cm,height=15cm]{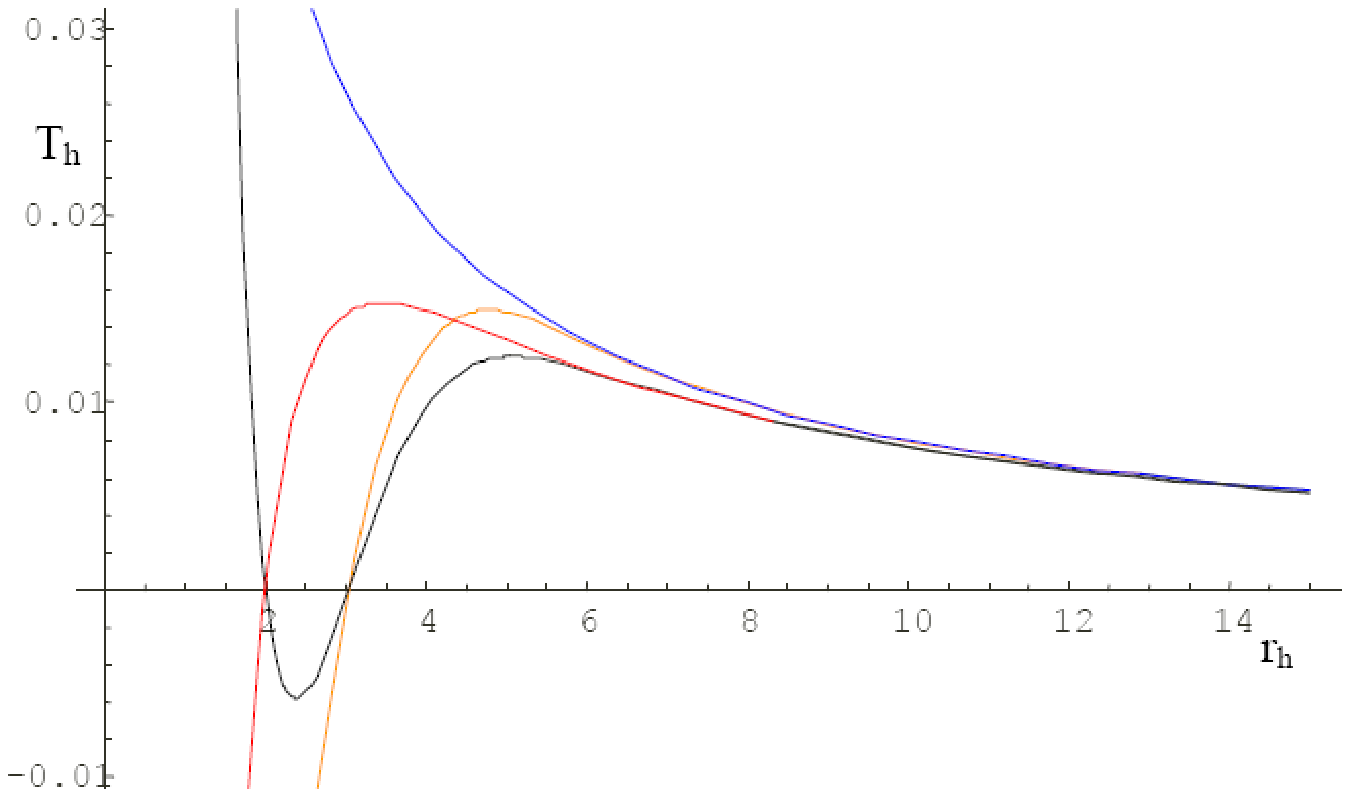}
\caption[]{{\it{$T_h$ Vs. $r_h$ plot (Here $|\alpha|=\theta$,
$\alpha$
is negative but $\theta$ is positive).\\
$r_h$ is plotted in units of $\sqrt\theta$ and $T_h$ is plotted in
units of $\frac{1}{\sqrt\theta}$.\\
Red curve: $\alpha\neq 0, \theta=0$.\\
Blue curve: $\alpha= 0, \theta= 0$.\\
Black curve: $\alpha\neq 0, \theta\neq 0$.\\
Yellow curve: $\alpha= 0, \theta\neq 0$.}}} \label{fig3}
\end{figure}

One problem with (\ref{4.61}) is that as $M$ approaches zero due
to Hawking radiation $T_h$ diverges. This is the Hawking paradox.
We shall study the effect of noncommutativity and back reaction on
this paradox by graphical methods when the black hole horizon is
comparable to $\sqrt{\theta}$. The dependence of the temperature
$T_h$ on the horizon radius $r_h$ is plotted in fig.(\ref{fig2})
(with positive $\alpha$) and in fig.(\ref{fig3}) (with negative
$\alpha$).

Fig.(\ref{fig2}) shows that the effect of back reaction does not
change the qualitative nature of the graphs for both commutative
and noncommutative cases. Though all curves marge at large $r_h$,
in the region $r_h\simeq\sqrt\theta$ the noncommutative effect is
quite significant. The commutative curves diverge as
$r_h\rightarrow 0$ whereas the noncommutative curves (with or
without back reaction) are no longer of hyperbolic type, instead
they have a peak at $r_h\simeq 4.7\sqrt\theta$ and then fall
quickly to zero at $r_0=3\sqrt{\theta}$. This $r_0$ corresponds to
the radius of the extremal black hole. In the region $r_h<r_0$ the
temperature is negative which means there is no black hole. In
this way Hawking paradox can be avoided by means of introducing
noncommutativity. This point was first noted in \cite{nicolini}.

In fig.(\ref{fig3}) we see that, for  $\alpha\neq 0, \theta=0$ (red curve) $T_h$
becomes zero at $r_h=r_0\simeq 2.0\sqrt\theta$ . This together with the case $\alpha=0,\theta\neq0$ (yellow curve of (\ref{fig2}) and (\ref{fig3}))
therefore avoid the Hawking paradox. But when $\alpha$ and $\theta$
are both nonzero $T_h$ vanishes for $r_h\simeq 3.0\sqrt\theta$ and
$r_h=2.0\sqrt\theta$ and then it diverges. Since this is not
meaningful physically, $\alpha$ must be positive for the cases
where both noncommutativity and back reaction effects are present.

{\it{Entropy and the area law}}: Having obtained the Hawking
temperature of the black hole we calculate the Bekenstein-Hawking
entropy. The expression of entropy can be obtained from the second
law of thermodynamics. Using (\ref{2.7}) the temperature
(\ref{4.66}) can be approximately expressed in terms of $M$. To
the leading order, we obtain,
\begin{eqnarray}
T_h=\frac{M^2+\alpha}{8\pi
M^3}\Big[1-\frac{4M^5}{(M^2+\alpha)\theta\sqrt{\pi\theta}}e^{-\frac{M^2}
{\theta}}\Big]+{\cal{O}}(\frac{1}{\sqrt\theta}e^{-\frac{M^2}{\theta}})
\end{eqnarray}
Using the second law of thermodynamics
$dS_{{\textrm{bh}}}=\frac{dM}{T_{h}}$
the Bekenstein--Hawking entropy is found to be
\begin{eqnarray}
S_{{\textrm{bh}}}&\simeq& 4\pi
M^2-4\pi\alpha\ln{(\frac{M^2}{\alpha}+1)} \nonumber
\\
&-& 16\sqrt{\frac{\pi}{\theta}}M^3
e^{-\frac{M^2}{\theta}}+\textrm{const.(independent of $M$)}
\label{4.13}
\end{eqnarray}
In order to see the connection between entropy and the horizon
area we need to write the right hand side of above equation in
terms of the area. But the effect of back reaction on a metric is
unknown. So we take $\alpha\rightarrow 0$ limit in the above
equation to get
\begin{eqnarray}
S_{\textrm{bh}}=\int{\frac{dM}{T_h}}\simeq 4\pi
M^2-16M^3\sqrt{\frac{\pi}{\theta}}e^{-\frac{M^2}{\theta}}.
\label{1.11}
\end{eqnarray}
The same expression of Bekenstein-Hawking entropy can also be
obtained by the tunnelling method as shown earlier in \cite{BB} .
Now using (\ref{2.7}) we obtain the noncommutative horizon area
($A$),
\begin{eqnarray}
A = 4\pi r^2_h=16\pi
M^2-64\sqrt{\frac{\pi}{\theta}}M^3e^{-\frac{M^{2}}{\theta}}.
\label{1.12}
\end{eqnarray}
Comparing equations (\ref{1.11}) and (\ref{1.12}) we find that at
the leading order, the noncommutative black hole entropy satisfies
the area law,
\begin{eqnarray}
S_{\textrm{bh}}=\frac{A}{4} \label{1.13}
\end{eqnarray}
which is functionally identical to the Benkenstein-Hawking area
law in the commutative space. This naturally arises the question
whether this law is true up to all orders in $\theta$ or not. To
address this issue we follow a method based on graphical analysis.
First we write the second law of thermodynamics in terms of the
horizon $r_h$ of the black hole. Using (\ref{2.6}) and
(\ref{1.08}) we get close form relation
\begin{eqnarray}
\frac{dS_{{\textrm{bh}}}}{dr_h}=\frac{\pi^{\frac{3}{2}}r_h}{\gamma(\frac{3}{2},
\frac{r^2_h}{4\theta})}. \label{1.3}
\end{eqnarray}
On the other hand (\ref{1.12}) and (\ref{1.13}) yield,
\begin{eqnarray}
\frac{dS_{{\textrm{bh}}}}{dr_h}\Big|_{\textrm{semiclassical}}=
\frac{dS_{{\textrm{bh}}}}{dA}\frac{dA}{dr_h}=2\pi r_h. \label{1.4}
\end{eqnarray}
Though in the large $r_h$ limit (\ref{1.3}) reduces to (\ref{1.4})
when $r_h\simeq\sqrt{\theta}$ there is a mismatch. In fact we need
a entropy--area law which is correct in the region $r_h\ge
3.0\sqrt{\theta}$, because when $r_h$ is smaller than that black
hole does not exist.

To find the corrections to the semiclassical area law we rewrite
(\ref{1.3}) in the following way
 \begin{eqnarray}
\frac{dS_{{\textrm{bh}}}}{dr_h}&=&\frac{\pi^{\frac{3}{2}}r_h}
{\frac{\sqrt\pi}{2}-\Gamma(\frac{3}{2},\frac{r^2_h}{4\theta})}
\nonumber\\
&=&2\pi
r_h\Big[1+\frac{2}{\sqrt\pi}\Gamma(\frac{3}{2},\frac{r^2_h}
{4\theta})+\frac{4}{\pi}\Gamma^2(\frac{3}{2},\frac{r^2_h}{4\theta})+........\Big].
\label{2.2}
\end{eqnarray}
The claim is that the terms involving the upper incomplete gamma
functions are the corrections to the area law. The justification
is given by the help of graphical analysis.
\begin{figure}[i]
\centering
\includegraphics[angle=0,width=15cm,height=15cm]{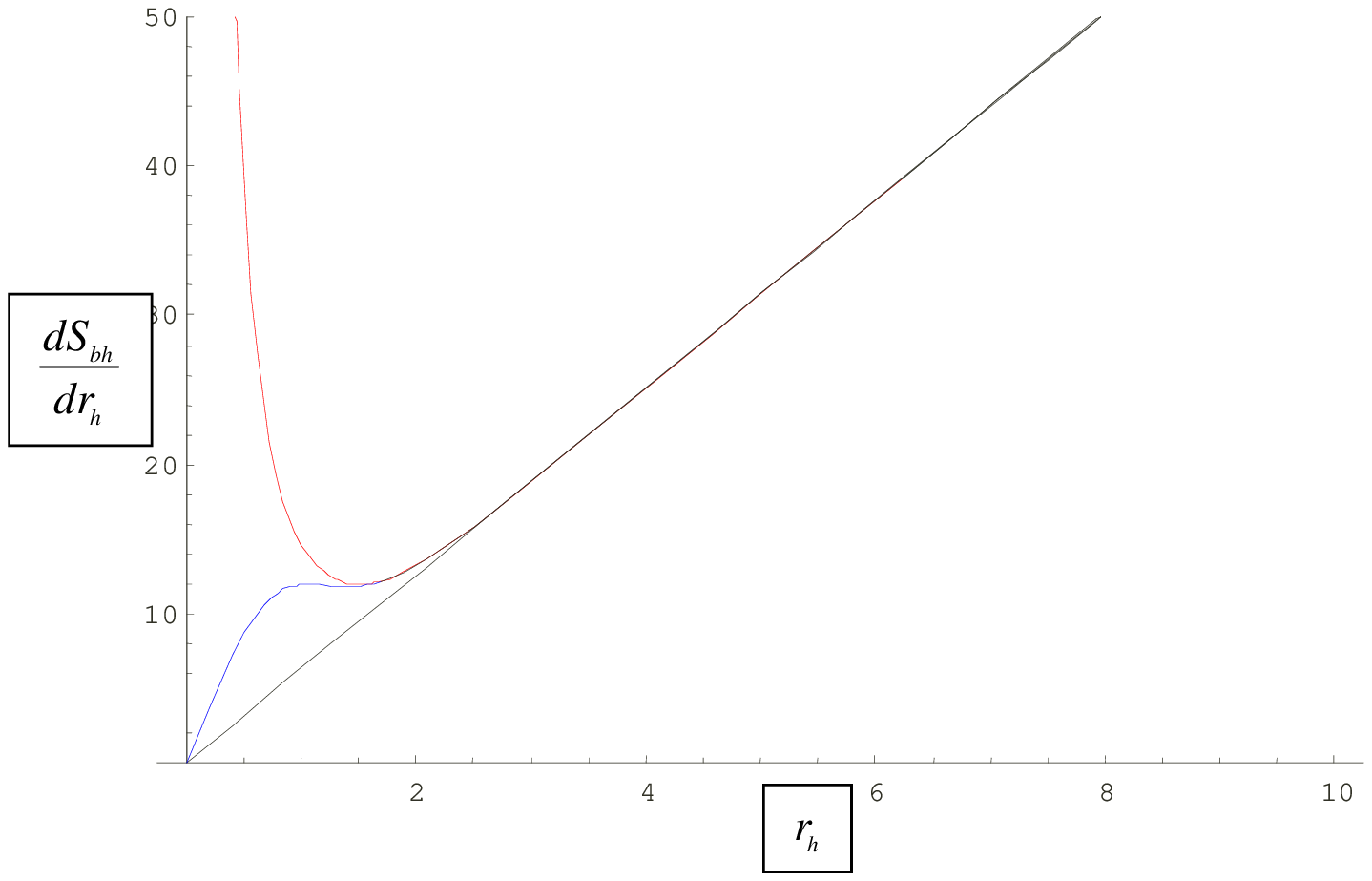}
\caption[]{{\it{$\frac{dS_{\textrm{bh}}}{dr_h}$ Vs. $r_h$ plot.\\
$\frac{dS_{\textrm{bh}}}{dr_h}$ is plotted in units of $4\theta$
and $r_h$
is plotted in units of $2\sqrt\theta$\\
Red curve: for eq. (\ref{1.3})\\
Black curve: for eq. (\ref{1.4})\\
Blue curve: for eq. (\ref{2.2}) (up to $\Gamma^2$)}}.}
\label{fig4}
\end{figure}
The graph of the above equation up to $\Gamma^2$ term is given in
figure (\ref{fig4}) along with equations (\ref{1.3}) and
(\ref{1.4}). It shows that the blue curve coincides with the red
curve for the entire physical domain $r_h\geq3.0\sqrt\theta$.
Inclusion of the third order correction improves the situation
even better\cite{B}. Thus we integrate (\ref{2.2}) over $r_h$ to
obtain
\begin{eqnarray}
S_{{\textrm{bh}}}&=& \pi r^2_h-\sqrt{\frac{\pi}{\theta}}~r^3_h
e^{-\frac{r^2_h}{4\theta}}-\sqrt{\pi\theta}~r_h
e^{-\frac{r^2_h}{4\theta}}-6\pi\theta\Big(1-{\textrm
{Erf}}(\frac{r_h}{2\sqrt\theta})\Big) \nonumber
\\
&+&2\sqrt\pi ~r^2_h
\Gamma(\frac{3}{2},\frac{r^2_h}{4\theta})+8\int
r_h\Gamma^2(\frac{3}{2},\frac{r^2_h}{4\theta})dr_h. \label{2.5}
\end{eqnarray}
Making use of (\ref{1.12}), the above equation is finally written
in terms of the noncommutative area as,
\begin{eqnarray}
S_{{\textrm{bh}}}&=&
\frac{A}{4}-\frac{A^{\frac{3}{2}}}{8\pi\sqrt\theta}
e^{-\frac{A}{16\pi\theta}}-\frac{\sqrt{\theta A}}{2}
e^{-\frac{A}{16\pi\theta}}-6\pi\theta\Big(1-{\textrm
{Erf}}(\frac{1}{4}\sqrt\frac{A}{\pi\theta})\Big) \nonumber
\\
&+&\frac{A}{2\sqrt\pi}
\Gamma(\frac{3}{2},\frac{A}{16\pi\theta})+\frac{1}{\pi}\int
\Gamma^2(\frac{3}{2},\frac{A}{16\pi\theta})dA. \label{2.51}
\end{eqnarray}
The first term is the noncommutative version of the
Bekenstein-Hawking area law whereas the other terms are the
different order corrections which contain exponentials of the
noncommutative semiclassical area $A$ and the error function.
Expectedly, in the limit $\theta\rightarrow 0$, all correction
terms vanish and one recovers the celebrated Bekenstein--Hawking
area law.

\section{Concluding Remarks:}
In the present article we have tried to convey the flavors of
noncommutative geometry inspired physics that have created a lot
of interest in recent times. Indeed the study is by no means
exhaustive simply because the topic is too vast and varied to be
covered meaningfully in a single article. Nevertheless this study
is representative, covering those aspects which have both physical
and mathematical interest.

From the physical perspective the effect of change of statistics
in many body systems presents an intriguing situation. Likewise,
as this review shows, Doubly Special Relativity acts as a link
between noncommutativity and quantum gravity. Noncommutativity
also prevents the singularity problem of Schwarzschild black hole
by preventing its absolute collapse. The presence of
noncommutativity naturally acts like a de-Sitter core beyond which
no further collapse takes place.

From the mathematical point of view a detailed study of twisted
gauge symmetries was given. This is normally left out in standard
articles on twists which concentrate basically on twisted Poincare
symmetry. The present discussion therefore compliments existing
analysis. Moreover it was reassuring to observe that approaches
used in conventional gauge theories with appropriate modifications
were revealed even in the noncommutative context. Although the
specific topics we have  covered are far from being completely
understood and we hope that this article will provide the
necessary impetus for further research.
\vskip .2cm
\section*{Acknowledgments} We thank Shailesh Kulkarni and Sudipta Das for their help.

\newpage


\baselineskip=1.6pt
\begin{thebibliography}{99}
\bibitem{szabo} M.R.Douglas and N.A.Nekrasov, Rev. Mod. Phys.
{\bf 73} 977 (2001) [hep-th/0106048]; R.~J.~Szabo, Phys. Rep. {\bf 378} 207 (2003) [hep-th/0109162].
\bibitem{sn} H.~S.~Snyder,
Phys. Rev. {\bf 71}  38 (1947); {\it{ibid}} {\bf 72}  68 (1947). 
\bibitem{yang} C. N. Yang, Phys. Rev. {\bf 72} 874 (1947) .
\bibitem{sw} N.~Seiberg, E.~Witten, JHEP {\bf 9909} 032 (1999)
[hep-th/9908142].
\bibitem{shei} M. M. Sheikh-Jabbari, Phys. Lett. {\bf B 455}, 129 (1999)  [hep-th/9901080];
V. Schomerus, 
  JHEP {\bf 06} 030 (1999)  [hep-th/9903205];
P. M. Ho and Y.-S. Wu,  Phys. Lett. {\bf B 398} 52 (1997)  [hep-th/9611233];
{\it{ibid}} Phys. Rev. {\bf D 58}, 066003 (1998) [hep-th/9801147].
\bibitem{biswa1} R. Banerjee, B. Chakraborty and S.Ghosh, Phys. Lett.
{\bf B 537} 340 (2002) [hep-th/0203199]; R. Banerjee, B. Chakraborty and K. Kumar, Nucl. Phys.
{\bf B 668} 179 (2003) (hep-th/0306122); B. Chakraborty, S. Gangopadhyay, A. G. Hazra and
F. G. Scholtz, Phys.Lett. {\bf B 625}  302 (2005) [hep-th/0508156]; C. Chatterjee, S. Gangopadhyay, A. Ghosh Hazra, S. Samanta 
 Int. J. Theor. Phys. {\bf 47} 2372 (2008)
 [arXiv:0801.4189].
\bibitem{biswa2} B. Chakraborty, S. Gangopadhyay and A. G. Hazra,
Phys. Rev. {\bf D 74} 105011 (2006) [hep-th/0608065].
\bibitem{ncsol}  R. Gopakumar, S. Minwalla and A. Strominger, JHEP {\bf 0005} 020 (2000) [hep-th/0003160];
 K. Dasgupta, S. Mukhi and G. Rajesh,  JHEP {\bf 0006} 022 (2000) [hep-th/0005006];
     J. A. Harvey, P. Kraus, F. Larsen and E. J. Martinec, JHEP {\bf 0007} 042 (2000) [hep-th/0005031]; M. Hamanaka, hep-th/0303256.
\bibitem{nccs} V. P. Nair and A. P. Polychronakos,  Phys. Rev. Lett. {\bf 87}  030403 (2001) [hep-th/0102181];
J. A. Harvey, hep-th/0105242.
\bibitem{ban-yang}  R. Banerjee, H. Seok Yang
     Nucl. Phys. {\bf B 708}  434 (2005) [hep-th/0404064];   V. O. Rivelles
     Phys. Lett. {\bf B 558} 191 (2003) [hep-th/0212262].
\bibitem{ban-yangnew} V. O. Rivelles, hep-th/0305122 ; E. Harikumar, V. O. Rivelles
    Class. Quant. Grav. {\bf 23} 7551 (2006) [hep-th/0607115;  P. Aschieri, M. Dimitrijevic, F. Meyer, J. Wess,
     Class. Quant. Grav. {\bf 23} 1883 (2006) [hep-th/0510059]; 
P. Aschieri, C. Blohmann, M. Dimitrijevic, F. Meyer, P. Schupp, J. Wess,
     Class. Quant. Grav. {\bf 22}  3511 (2005) [hep-th/0504183].
\bibitem{anomaly} J. M. Gracia-Bondia and C. P. Martin, Phys. Lett. {\bf B 479} 321 (2000) [hep-th/0002171];
L. Bonora, M. Schnabl and A. Tomasiello, Phys. Lett. {\bf B 485} 311 (2000) [hep-th/0002210];
F. Ardalan and N. Sadooghi, Int. J. Mod. Phys. {\bf A 16} 3151 (2001) [hep-th/0002143];
C. P. Martin, Nucl. Phys. {\bf B 623}150 (2002) [hep-th/0110046]; R. Banerjee and S. Ghosh
Phys. Lett. {\bf B 533} 162 (2002) [hep-th/0110177]; R. Banerjee, K. Kumar
     Phys. Rev. {\bf D 75} 045008 (2007) [hep-th/0604162 ]; {\it{ibid}}, Phys. Rev. {\bf D 72} 085012 (2005) [hep-th/0505245]; {\it{ibid}}, Phys. Rev. {\bf D 71}  045013 (2005) [hep-th/0404110]; 
   R. Banerjee, C. Lee, H. S. Yang
     Phys. Rev. {\bf D 70} 065015 (2004) [arXiv:hep-th/0312103];  
     R. Banerjee
     Int. J. Mod. Phys. {\bf A 19}  613 (2004) [hep-th/0301174].
\bibitem{ncsol1} S. Ghosh, Nucl. Phys. {\bf B 670} 359 (2003) [hep-th/0306045]; {\it{ibid}} Phys. Rev. {\bf D 70} 085007 (2004) [hep-th/0402029]; {\it{ibid}} Annals. Phys. {\bf 318} 432 (2005); T. R. Govindarajan, E. Harikumar, Phys.Lett. {\bf B 602} 238 (2004) [hep-th/0406273]; B. Chakraborty  , S. Ghosh and R. P. Malik, Nucl. Phys. {\bf B 600} 351 (2001) [hep-th/0008168].
\bibitem{nccsg} S. Ghosh, Phys. Lett. {\bf B 563} 112 (2003) [hep-th/0303022];{\it{ibid}} Phys. Lett. {\bf B 558} 245 (2003) [hep-th/0210107], Mod. Phys. Lett. {\bf A 20} 1227 (2005) [hep-th/0407086];
     M. Botta Cantcheff and P.Minces, Phys. Lett. {\bf B 557} 283 (2003) [hep-th/0212031]; O. F. Dayi,
     Phys. Lett. {\bf B 560} 239 (2003) [hep-th/0302074]; M. Botta Cantcheff and P. Minces, Eur. Phys. J.  {\bf C 34} 393 (2004) [hep-th/0306206]; M. S. Guimaraes, D. C.Rodrigues, C. Wotzasek and J. L. Noronha, Phys. Lett. {\bf B 605} 419 (2005) [hep-th/0410156];  E. Harikumar, Victor O. Rivelles
         Phys. Lett. {\bf B 625} 156 (2005) [hep-th/0506078]; {\it{ibid}} hep-th/0610098.
\bibitem{exnc} R. Banerjee, S. Ghosh, T. Shreecharan, Phys. Lett. {\bf B 662} 231 (2008) [arXiv:0712.3631].
\bibitem{cpm} C. P. Martin, C. Tamarit, Phys. Lett. {\bf B 658} 170 (2008) [arXiv:0706.4052].
\bibitem{ncho} V. P. Nair and A. P. Polychronakos, Phys. Lett. {\bf B 505} 267 (2001) [hep-th/0011172]; J. Gamboa, M. Loewe, and J. C. Rojas,  Phys. Rev. {\bf D 64} 067901 (2001) [hep-th/0010220]; S. Bellucci, A. Nersessian, and C. Sochichiu,  Phys. Lett. {\bf B 522} 345 (2001) [hep-th/0106138]; R. Banerjee,
     Mod. Phys. Lett. {\bf A 17} 631 (2002) (hep-th/0106280) ; B. Muthukumar and P. Mitra, Phys. Rev. {\bf D 66} 027701 (2002) [hep-th/0204149]; S. Samanta
    Mod. Phys. Lett. {\bf A 21} 675 (2006) [hep-th/0510138] ;  S. Gangopadhyay, F. G. Scholtz, arXiv:0812.3474.
\bibitem{nch}    M. Chaichian, M. M. Sheikh-Jabbari and A. Tureanu, Phys. Rev. Lett. 86, 2716 (2001) [arXiv:hep-th/0010175];  X. Calmet, Eur.Phys.J.C41:269 (2005) [arXiv:hep-ph/0401097];  Z. Guralnik, R. Jackiw, S.-Y. Pi and A. P. Polychronakos, Phys. Lett. B517, 450 (2001) [arXiv:hep-th/0106044].
\bibitem{bertol}O. Bertolami, J. G. Rosa, C. M. L. de Aragao, P.Castorina and D.
Zappala,  Phys. Rev. D
72, 025010 (2005) (hep-th/0505064); R. Banerjee, B. Dutta Roy, S. Samanta
      Phys.Rev.D74:045015,2006 (hep-th/0605277); A. Saha, arXiv:0803.3957; S. Samanta, arXiv:0804.0172. 
  \bibitem{uvir} T. Filk, Phys. Lett. {\bf B 376} 53 (1996);
 N. Ishibashi, S. Iso, H. Kawai and Y. Kitazawa, Nucl. Phys. {\bf B 573} 573 (2000) [hep-th/9910004];
 S. Minwalla, M. Van Raamsdonk and N. Seiberg, JHEP {\bf 0002} 020 (2000) [hep-th/9910004].    
\bibitem{wul} H. Grosse and R. Wulkenhaar, Commun. Math. Phys. {\bf 256} 305 (2005) [hep-th/0305066]; E. Langmann, R. J. Szabo and K. Zarembo, JHEP {\bf 0401} 017 (2004) [hep-th/0308043]; H. Grosse and R. Wulkenhaar, Eur. Phys. J. {\bf C 35} 277282 (2004) [hep-th/0402093]; V. Rivasseau, {\it{Seminaire Poincare X}}, 2007.
\bibitem{con} A. Connes and M. A. Rieffel,  Contemp.
    Math. {\bf 62} 237 (1987) ; A. Connes and J. Lott,  Nucl. Phys.
    (Proc. Suppl.) {\bf B 18} 29 (1990);
    C.P. Martin, ́ J. M. Gracia-Bond ́ and J. C. V ́rilly,  Phys. Rep. {\bf 294} 363 (1998)
    [hep-th/9605001];   Ali H. Chamseddine, A. Connes
     Phys. Rev. Lett. {\bf 99}191601 (2007) [arXiv:0706.3690]; A. Connes,
     JHEP {\bf 0611} 081 (2006) [arXiv:hep-th/0608226]; A. Connes, {\it{Seminaire Poincare X}}, 2007.
\bibitem{fluid} L. Susskind, arXiv: hep-th/0101029; A. P. Polychronakos, JHEP {\bf 0104} 011 (2001) [hep-th/0103013]; A. P. Polychronakos, {\it{Seminaire Poincare X}}, 2007.
\bibitem{horv} C. Duval, P. A. Horvathy, Phys. Lett. {\bf B 594} 402 (2004) [arXiv:hep-th/0402191];
V. P. Nair and A. P. Polychronakos, Phys. Lett. {\bf B 505} 267 (2001) [hep-th/0011172].
\bibitem{cnp} M. S. Plyushchay, Phys. Lett. {\bf B 248} 107 (1990);  P. A. Horvathy, M. S. Plyushchay, Phys. Lett. {\bf B 595} 547 (2004); 
P. A. Horvathy, M. S. Plyushchay, Nucl. Phys. {\bf B 714} 269 (2005); R. Jackiw and V. P. Nair, Phys. Rev {\bf D 43} 1933 (1991);  C. Chou, V. P. Nair and A. P. Polychronakos, Phys. Lett. {\bf B 304} 105 (1993) [hep-th/9301037]; S. Ghosh,  Phys. Lett. {\bf B 338} 235 (1994),
 Erratum-ibid. {\bf B 347} 468(1995) [hep-th/9406089];  
Gerald V. Dunne, R. Jackiw, C. A. Trugenberger, Phys. Rev. {\bf D 41} 661 (1990).
\bibitem{scholtz} F. G. Scholtz, B. Chakraborty, S. Gangopadhyay, J. Govaerts
     J. Phys. {\bf A 38} 9849 (2005) [cond-mat/0509331].
\bibitem{ber} A. Berard and H. Mohrbach, Phys. Rev. {\bf D 69} 127701 (2004) [hep-th/0310167]; P. Gosselin, A. Bérard, H. Mohrbach,
Eur. Phys. J. {\bf B 58} 137 (2007) [hep-th/0603192]; K. Yu. Bliokh and Yu. P. Bliokh, Annals. Phys. {\bf 319} 13 (2005) [quant-ph/0404144].
\bibitem{ahe} T. Jungwirth, Q. Niu and A. H. M acdonald, Phys. Rev. Lett. {\bf 88} 207208 (2002); M. Onoda and N. Nagaosa, Phys. Rev. Lett. {\bf 93} 083901 (2004); D. Xiao, J. Shi and Q. Niu, Phys. Rev. Lett. {\bf 95} 137204 (2005) [cond-mat/0502340];  C. Duval, Z. Horvath, P. A. Horvathy and L. Martina,  Phys. Rev. Lett. {\bf 96} 099701 (2006) [cond-mat/0509806].
\bibitem{she} S. Murakami, N. Nagaosa and S. C. Zhang, Science {\bf 301} 1348 (2003);
 J. Sinova, D. Culcer, Q. Niu, N. A. Sinitsyn, T. Jungwirth and A. H. Macdonald., Phys. Rev. Lett. {\bf 92} 126603 (2004); S. Dhar, B. Basu and S. Ghosh,  Phys. Lett. {\bf A 371} 406  (2007) [cond-mat/0701096]
\bibitem{ber2} P. Gosselin, A. Bérard, H. Mohrbach and S. Ghosh, Eur. Phys. J. {\bf C 59} 883 (2009) [arXiv:0802.3565].
\bibitem{jmm} R. Jackiw, Phys. Rev. Lett. {\bf 54} 159 (1985); {\it{ibid}} Int. J. Mod. Phys. {\bf A19S1} 137 (2004) [hep-th/0212058].
\bibitem{chm} M. Chaichian, S.  Ghosh, M. Langvik and A. Tureanu, Phys. Rev. {\bf D 79} 125029 (2009) [arXiv:0902.2453].
\bibitem{4} B. Chakraborty, S. Gangopadhyay, A. G. Hazra and F. G. Scholtz, J. Phys. A {\bf 39} 9557 (2006) [hep-th/0601121].
\bibitem{5} E. Akofor, A. P. Balachandran, A. Joseph, Int. J. Mod. Phys. {\bf A 23} 1637 (2008) [arXiv:0803.4351].
\bibitem{6} A. P. Balachandran, G. Mangano, A. Pinzul and S. Vaidya, Int. J. Mod. Phys. {\bf A 21} 3111 (2006) [hep-th/0508002].
\bibitem{gho2} S.~Ghosh and  P.~Pal, Phys.~Rev.~{\bf D 75} 105021 (2007) [hep-th/0702159].
\bibitem{am} G. Amelino-Camelia, Nature {\bf 418} 34 (2002) [gr-qc/0207049]; {\it{ibid}} Phys. Lett. {\bf B 510} 255 (2001) [hep-th/0012238]; {\it{ibid}} Int. J. Mod. Phys. {\bf D11}
35 (2002) [gr-qc/0012051].
\bibitem{rb} R. Banerjee and S. Samanta, JHEP {\bf 0702} 046 (2007) [hep-th/0611249].
\bibitem{samanta1} R. Banerjee and S. Samanta
Eur. Phys. J.  {\bf C 51} 207 (2007) [arXiv: hep-th/0608214].
\bibitem{jWess} J. Wess, J. Phys. Conf. Ser. {\bf 53} 752 (2006)
[hep-th/0608135]; {\it{ibid}} hep-th/0408080.  
\bibitem{cht} M. Chaichian and A. Tureanu, Phys. Lett. {\bf B 637} 99 (2006) [hep-th/0604025];
\bibitem{cht1} M. Chaichian, P. Prešnajder, A. Tureanu, Phys. Rev. Lett. {\bf 94} 151602 (2005)  [hep-th/0409096]; M. Chaichian, P. Kulish, K. Nishijima, A. Tureanu, Phys. Lett. {\bf B 604} 98 (2004) [hep-th/0408069].
\bibitem{l1}    R. Banerjee, B. Chakraborty, K. Kumar
     Phys. Rev. {\bf D 70} 125004 (2004) [hep-th/0408197]; S. Ghosh, hep-th/0310155.
\bibitem{MS} P. Mukherjee and A. Saha, Phys. Rev. {\bf{ D 74}} 027702 (2006) [hep-th/0605287].
\bibitem{CTZ} M. Chaichian, A. Tureanu and G. Zet, Phys. Lett  {\bf{B 660}} 573 (2008) [arXiv:0710.2075].
\bibitem{glauber} R. J. Glauber, Phys. Rev. {\bf 131} 2766 (1963)
\bibitem{nicolini} P. Nicolini, A. Smailagic and E. Spallucci,  Phys. Lett. {\bf B 547} 632 (2006) [gr-qc/0510112].
\bibitem{kupr} V. G. Kupriyanov, D. V. Vassilevich, Eur. Phys. J. {\bf C 58} 627 (2008).
\bibitem{2} N. Macris and S. Ouvry, J. Phys. A {\bf 35} 4477 (2002) [hep-th/0112181].
\bibitem{newref} F. G. Scholtz, B. Chakraborty, S. Gangopadhyay, A. Ghosh Hazra, Phys. Rev. {\bf D 71} 085005 (2005) [hep-th/0502143].
\bibitem{7} M. Chaichian, P. Kulish, K. Nishijima and A.Tureanu, Phys. Lett. {\bf B 604} 98 (2004) [hep-th/0408069].
\bibitem{wess} M. Dimitrijevic, L. Jonke, L. Moller, E. Tsouchnika, J. wess , Eur. Phys. J. {\bf C 31} 129 (2003) [hep-th/0307149].
\bibitem{8} S. Khan, B. Chakraborty and F. G. Scholtz, Phys. Rev.  {\bf D 78}  025024 (2008) [arXiv:0707.4410].
\bibitem{topan} P. G. Castro, B. Chakraborty and F. Toppan,
J. Math. Phys. {\bf 49} 082106 (2008) [arXiv:0804.2936].
\bibitem{fiore} G. Fiore and J. Wess, Phys. Rev. {\bf{D 75}} 105022 (2007).
\bibitem{pi} G. Piacitelli, arxiv:0902.0575.
 \bibitem{balanewref} A. P. Balachandran, T. R. Govindarajan, G. Mangano, A. Pinzul, B. Qureshi, S. Vaidya, Phys. Rev. {\bf D 75} 045009 (2007) [hep-th/0608179]; A. Tureanu, Phys. Lett. {\bf B 638} 296 (2006) [hep-th/0603219].
\bibitem{9} F. G. Scholtz, B. Chakraborty, J. Govaerts and S.Vaidya, J. Phys. A 1458 (2007) [arXiv:0709.3357].
\bibitem{sunan} S. Gangopadhyay and F. G. Scholtz, Phys. Rev. Lett. {\bf{102}} 241602 (2009) [arxiv:0904.0379].
\bibitem{cite1} C. Duval, P. A. Horvathy,  Phys. Lett. {\bf B 479} 284 (2000). 
\bibitem{cite2} V. P. Nair, A. P. Polychronakos,  Phys. Lett. {\bf B 505} 267 (2001).
\bibitem{10} F. G. Scholtz, L. Gouba, A. Hafver, C. Rohwer, J. Phys. {\bf A 42}175303 (2009) [arXiv:0812.2803].
\bibitem{11} F. G. Scholtz, J. Govaerts, J. Phys.  {\bf A 41} 505003 (2008) [arXiv:0810.3064].
\bibitem{planck} C. Rovelli and L. Smolin, Nucl. Phys. {\bf B 442} 593 (1995); Erratum: {\it{ibid}} {\bf 456} 734 (1995) [C. Rovelli and L. Smolin].
\bibitem{mdis}
R.~Gambini and J.~Pullin, Phys.~Rev. {\bf D  59} 124021 (1999) [gr-qc/9809038];
T.~Kifune,; Astrophys.~J.~Lett.~ 518, L21 (1999) [astro-ph/9904164]; J.~Alfaro,
H.A.~Morales-Tecotl and L.F.~Urrutia, Phys.~Rev.~Lett. {\bf 84} 2318 (2000) [gr-qc/9909079]; S. D.~Biller et al.,
Phys.~Rev.~Lett.~ {\bf 83} 2108 (1999) [gr-qc/9810044]; J.P.~Norris, J.T.~Bonnell,
G. F.~Marani and J. D.~Scargle, astro-ph/9912136; A.~de Angelis,
astro-ph/0009271;
 R.~Aloisio, P.~Blasi, P.L.~Ghia and A.F.~Grillo,
 Phys.~Rev.~{\bf D 62} 053010 (2000) [astro-ph/0001258];
 R.J.~Protheroe and H.~Meyer, Phys.~Lett.~{\bf B 493} 1 (2000) [astro-ph/0005349].
 H.~Sato, astro-ph/0005218.
\bibitem {mag} J. Magueijo and L. Smolin, Phys. Rev. Lett. {\bf 88} 190403 (2002) [hep-th/0112090]; {\it{ibid}}
Phys.Rev. {\bf D 67} 044017 (2003) [gr-qc/0207085].
\bibitem{dsr2} J. Lukierski, A. Nowicki, H. Ruegg and V. N. Tolstoy,
Phys. Lett. {\bf B264} 331 (1991); S. Majid and H. Ruegg, Phys. Lett. {\bf B 334}
348 (1994) [hep-th/9405107]; J. Lukierski, H. Ruegg, W. J. Zakrzewski ,
 Annals. Phys. 243 90 (1995) [hep-th/9312153].
\bibitem{dfr} S. Doplicher, K. Fredenhagen and J. E. Roberts, Phys. Lett. {\bf B 331} 39 (1994).
\bibitem{rev} J. Kowalski-Glikman, Lect. Notes Phys. {\bf 669}131 (2005) [hep-th/0405273].
\bibitem{kpon}  J.~Lukierski, A.~Nowicki and
H.~Ruegg, Phys.~Lett.~{\bf B 293} 344 (1992);  J.~Lukierski, H.~Ruegg and W.~Ruhl,
Phys.~Lett.~ {\bf B 313} 357 (1993).
\bibitem{kow} J. Kowalski-Glikman and S. Nowak, Phys. Lett. {\bf B 539} 126 (2002) [hep-th/0203040]; {\it{ibid}}
     Int. J. Mod. Phys. {\bf D12} 299 (2003) [hep-th/0204245]; {\it{ibid}} Class. Quant. Grav. {\bf 20} 4799 (2003) [hep-th/0304101]; J. M. Romero and
A. Zamora, Phys. Rev. {\bf D 70} 105006 (2004) [hep-th/0408193].
\bibitem{granik} A. Granik, hep-th/0207113; S. Mignemi, Phys. Rev. {\bf D 68}  065029 (2003)
 [gr-qc/0304029].
\bibitem{kimb} D. Kimberly, J. Magueijo and J. Medeiros, Phys. Rev. {\bf D 70} 084007 (2004) [gr-qc/0303067].
\bibitem{bru} N. R. Bruno, G.A melino-Camelia and J.Kowalski-Glikman,
Phys. Lett. {\bf B 522} 133 (2001) [hep-th/0107039].
\bibitem{fins} F. Girelli, S. Liberati and L. Sindoni, Phys. Rev. {\bf D 75} 064015 (2007) [gr-qc/0611024]; S. Hossenfelder, Phys. Lett. {\bf A 367} 11 (2007) [hep-th/0612167].
\bibitem{garcia} J. Antonio Garcia, Phys. Rev. {\bf D 76} 048501 (2007) [arXiv:0705.0143];
B. F. Rizzuti, arXiv:0710.3724.
\bibitem{mel} S. Meljanac and M. Stojic, Eur. Phys. J. {\bf C 47} 531 (2006) [hep-th/0605133]; S. Meljanac, S. Kresic-Juric and M. Stojic, Eur. Phys. J. {\bf C 51} 229 (2007) [hep-th/0702215].
\bibitem{goss} P.~Gosselin, A.~Berard, H.~Mohrbach and S.~Ghosh,
Phys. Lett. {\bf B 660} 267 (2008) [arXiv:0709.0579].
\bibitem{am1} A. Agostini, G. Amelino-Camelia and M. Arzano,
Class. Quant. Grav. {\bf 21} 2179 (2004) [gr-qc/0207003].
\bibitem{others} A. A. Deriglazov, JHEP {\bf 0303} 021 (2003) [hep-th/0211105]; A. Pinzul,
A. Stern, Phys. Lett. {\bf B 593} 279 (2004) [hep-th/0402220]; S. Ghosh,
Phys. Lett. {\bf B 648} 262 (2007) [hep-th/0602009]; {\it{ibid}} Phys. Rev. {\bf D 74} 
084019 (2006) [hep-th/0608206]; {\it{ibid}} Phys. Lett. {\bf B 623} 251 (2005) [hep-th/0506084];
S. Ghosh and P. Pal, Phys. Lett. {\bf B 618} 243 (2005) [hep-th/0502192]; F. Girelli,
T. Konopka, J. Kowalski-Glikman and E. R. Livine, Phys. Rev. {\bf D 73} 045009 (2006)
[hep-th/0512107]; L. Freidel, F. Girelli and E. R. Livine, Phys. Rev. {\bf D 75} 105016 (2007) [hep-th/0701113]; R. Banerjee, S. Kulkarni and S. Samanta, JHEP {\bf 0605} 077 (2006) 
[hep-th/0602151].
\bibitem{dirac} P. A. M.Dirac, {\it{Lectures on Quantum Mechanics}},
Yeshiva University Press, New York, 1964.
\bibitem{vass} D. V. Vassilevich, Mod. Phys. Lett. {\bf A 21} 1279 (2006) [arXiv:hep-th/0602185]
\bibitem{Asch} P. Aschieri, M. Dimitrijevic, F. Meyer, S. Schraml and J. Wess
Lett. Math. Phys. {\bf 78} 61 (2006) [arXiv:hep-th/0603024].
\bibitem{gitman} D. M. Gitman and I. V. Tyutin, Quantization of Fields with Constraints,
      Springer-Verlag Berlin, Heidelberg (1990).
\bibitem{shirzad} A. Shirzad, J.
      Phys. {\bf A 31} 2747 (1998).
\bibitem{rothe1} R. Banerjee, H. J. Rothe and K. D. Rothe, Phys. Lett. {\bf B 479} 429 (2000) [hep-th/9907217]; 
 arXiv:hep-th/9909039  R. Banerjee, H.J. Rothe, K.D. Rothe 
 J.Phys. {\bf{A33}} 2059 (2000). 
\bibitem{rothe2} R. Banerjee, H. J. Rothe and K. D. Rothe, Phys. Lett. {\bf B 463} 248 (1999) [hep-th/9906072].
\bibitem{Hennaux} M. Hennaux and C. Teitelboim, Quantization of Gauge Systems, Princeton University
Press, Princeton (1992).
\bibitem{ag} L.~Alvarez-Gaume, F.~Meyer, M.~A.~Vazquez-Mozo, Nucl. Phys. {\bf{ B 753}} 92 (2006) [hep-th/0605113].
\bibitem{samanta} S.  Samanta, arXiv:0708.3300.
\bibitem{amorim} R. Amorim and F. A. Farias, Phys. Rev.  {\bf D 65} 065009 (2002)[hep-th/0109146].
\bibitem{banerjee22} R. Banerje, Phys. Rev.  {\bf  D 67} 105002 (2003) [hep-th/0210259].
 
\bibitem{szabog}R.~J.~Szabo, Classical Quantum Gravity {\bf{23}} R199 (2006) [hep-th/0606233];  E. Langmann, R. J. Szabo, Phys. Rev. {\bf D 64} 104019 (2001) [hep-th/0105094].
\bibitem{Chamseddine:2000si}
  A.~H.~Chamseddine,  Phys.\ Lett.\  {\bf B 504} 33 (2001) [hep-th/0009153].
\bibitem{Aschieri:2005yw} P.~Aschieri, C.~Blohmann, M.~Dimitrijevic, F.~Meyer, P.~Schupp, J.~Wess
 Class. Quant. Grav. {\bf 22} 3511 (2005) [hep-th/0504183].
\bibitem{CK1} X.~Calmet, A.~Kobakhidze, Phys. Rev. {\bf D 72} 045010 (2005) [hep-th/0506157].
\bibitem{UNI}
J.~J.~van der Bij, H.~van Dam and Y.~J.~Ng, Physica {\bf 116A},
307 (1982); F.~Wilczek, Phys.\ Rept.\  {\bf 104}, 143 (1984);
W.~Buchmuller and N.~Dragon, Phys.\ Lett. {\bf B 207} 292
(1988); M.~Henneaux and C.~Teitelboim, Phys.\ Lett. {\bf B 222},
195 (1989); W.~G.~Unruh,
 Phys.\ Rev. {\bf D 40} 1048 (1989).
\bibitem{CK2} X.~Calmet, A.~Kobakhidze, Phys. Rev.  {\bf{ D 74}} 047702 (2006) [hep-th/0605275].
\bibitem{BMS} R Banerjee, P. Mukherjee, S. Samanta, Phys. Rev. {\bf{D 75}} 125020 (2007) [hep-th/0703128].
\bibitem{HR} E. Harikumar and Vivtor. O. Rivelles, Class. Quant. Grav. {\bf{23}} 7551 (2006) [hep-th/0607115].
\bibitem{K} A. Kobakhidze, arXiv:0712.0642 [gr-qc].
\bibitem{MS1} P. Mukherjee and A. Saha, Phys. Rev.  {\bf{D 77}} 064014 (2008) [arXiv:0710.5847].
\bibitem{CTZ1} M. Chaichian, M. R. Setare, A. Tureanu and G. Zet, JHEP {\bf{0804}} 064 (2008) [arXiv:0711.4546].
\bibitem{hsyang} M. Salizzoni, A. Torrielli, H. S. Yang, Phys. Lett. {\bf B 634} 427 (2006) [hep-th/0510249]; H. S. Yang, M. Salizzoni, Phys. Rev. Lett. {\bf 96} 201602 (2006) [hep-th/0512215]; H. S. Yang, hep-th/0608013; {\it ibid}, hep-th/0611174; {\it ibid}, Mod. Phys. Lett. {\bf A 22} 1119 (2007) [hep-th/0612231]; H. Steinacker, Nucl. Phys. {\bf B 810} 1 (2009) [arXiv:0806.2032]; D. Klammer, H. Steinacker, Phys. Rev. Lett. {\bf 102} 221301(2009) [arXiv:0903.0986].
\bibitem{nicrev} P. Nicolini, Int. J. Mod. Phys. {\bf A 24} 1229 (2009) [arXiv:0807.1939].
 \bibitem{grezia} E. D. Grezia, G. Esposito  and  G.Miele, Class. Quant. Grav. {\bf 23} 6425 (2006) [hep-th/0607157];
 K.Nozari and S.H.Mehdipour, Class. Quant. Grav. {\bf 25} 175015 (2008) [arXiv:0801.4074]. 
\bibitem{B} R. Banerjee, B. Ranjan Majhi, S. K. Modak, Class. Quant. Grav. {\bf 26} 085010 (2009) [arXiv:0802.2176]. 
\bibitem{madore} J. Madore, S. Schraml, P. Schupp, J. Wess
     Eur. Phys. J. {\bf C18} 78 (2001) [hep-th/0009230]; {\it{ibid}}  Eur. Phys. J. {\bf C16} 161 (2000) [hep-th/0001203].
 \bibitem{Jurco:2000ja}
B.~Jurco, S.~Schraml, P.~Schupp and J.~Wess, Eur.\ Phys.\ J.
{\bf  C 17} 521 (2000) [hep-th/0006246]. 
 \bibitem{chams} A. H. Chamseddine, Int. J. Mod. Phys. {\bf A 16} 759 (2001) [hep-th/0010268].
\bibitem{cite3} P. Nicolini,  J. Phys. {\bf A 38} L631 (2005). 
\bibitem{cite4} S. Ansoldi, P. Nicolini,  A. Smailagic, E. Spallucci,  Phys. Lett. {\bf B 645} 261 (2007). 
\bibitem{stern} S. Fabi, B. Harms, A. Stern  Phys. Rev.  {\bf{D 78}} 065037 (2008) [arXiv:0808.0943].
\bibitem{fro} V. P. Frolov, M. A. Markov and  V. F. Mukhanov Phys. Rev. {\bf D 41} 383 (1990).
\bibitem{dym} I. Dymnikova, Gen. Rel. Grav. {\bf 24} 235 (1992).   
 \bibitem{kim} Y. S. Myung, Yong-Wan Kim and Young-Jai Park, JHEP {\bf 02} 012 (2007) [gr-qc/0611130].
 \bibitem{BB} R. Banerjee, B. R. Majhi, S. Samanta, Phys. Rev. {\bf D 77}124035 (2008) [arXiv:0801.3583].
 \bibitem{Lousto} C. O. Lousto and N. Sanchez, Phys. Lett. {\bf B 212} 411 (1988).
 \bibitem{Fursaev} D. V. Fursaev, Phys. Rev. {\bf D 51} 5352 (1995) [hep-th/9412161].
 \bibitem{York} J. W. York, Jr., Phys. Rev. {\bf D 31} 755 (1985).
 \bibitem{majhi} R. Banerjee and B. R. Majhi, Phys. Lett. {\bf B 662} 62 (2008) [arXiv:0801.0200].
\end{thebibliography}
\end{document}